\documentclass[lettersize,journal]{IEEEtran}
\usepackage{amsmath,amsfonts}
\usepackage{array}
\usepackage[caption=false,font=normalsize,labelfont=sf,textfont=sf]{subfig}
\usepackage{textcomp}
\usepackage{stfloats}
\usepackage{url}
\usepackage{xcolor}
\usepackage{verbatim}
\usepackage{graphicx}
\usepackage{cite}
\usepackage{threeparttable}
\usepackage{booktabs}
\usepackage[numbers,sort&compress]{natbib}
\usepackage[ruled,vlined]{algorithm2e}

\SetCommentSty{mycommfont}
\setcitestyle{style=numeric-comp}

\graphicspath{{./fig/}}

\def\BibTeX{{\rm B\kern-.05em{\sc i\kern-.025em b}\kern-.08em
    T\kern-.1667em\lower.7ex\hbox{E}\kern-.125emX}}
\usepackage{balance}

\begin{document}

\title{Deep Learning-Based Modeling of 5G Core Control Plane for 5G Network Digital Twin}

\author{Zhenyu~Tao,
        Yongliang~Guo,
        Guanghui~He,
        Yongming~Huang,~\IEEEmembership{Senior Member,~IEEE}
        and~Xiaohu~You*,~\IEEEmembership{Fellow,~IEEE}
        
\thanks{Z.Tao is with the Southeast University, Nanjing, 211189, China (email: zhenyu\_tao@seu.edu.cn).}
\thanks{Y.Guo and G.He are with the Purple Mountain Laboratories, Nanjing, 211111, China (email: \{guoyongliang, heguanghui\}@pmlabs.com.cn).}
\thanks{Y.Huang and X.You are with the Southeast University, Nanjing, 211189, China, and the Purple Mountain Laboratories, Nanjing, 211111, China (email: \{huangym, xhyu\}@seu.edu.cn).}
\thanks{X.You is the corresponding author of this paper.}}

\markboth{IEEE TRANSACTIONS ON COGNITIVE COMMUNICATIONS AND NETWORKING, ~Vol.~15, No.~1, August~2023}%
{Shell \MakeLowercase{\textit{et al.}}: A Sample Article Using IEEEtran.cls for IEEE Journals}

\maketitle

\begin{abstract}
\textcolor{black}{Digital twin serves as a crucial facilitator in the advancement and implementation of emerging technologies within 5G and beyond networks. However, the intricate structure and diverse functionalities of the existing 5G core network, especially the control plane, present challenges in constructing core network digital twins. In this paper, we propose two novel data-driven architectures for modeling the 5G control plane and implement corresponding deep learning models, namely 5GC-Seq2Seq and 5GC-former, based on the Vanilla Seq2Seq model and Transformer decoder respectively. We also present a solution enabling the interconversion of signaling messages and length-limited vectors to construct a dataset. The experiments are based on 5G core network signaling messages collected by the Spirent C50 network tester, encompassing various procedures such as registration, handover, and PDU sessions. The results show that 5GC-Seq2Seq achieves a 99.997\% F1-score (a metric measuring the accuracy of positive samples) in single UE scenarios with a simple structure, but exhibits significantly reduced performance in handling concurrency. In contrast, 5GC-former surpasses 99.999\% F1-score while maintaining robust performance under concurrent UE scenarios by constructing a more complex and highly parallel model. These findings validate that our method accurately replicates the principal functionalities of the 5G core network control plane.}

\end{abstract}


%
\IEEEpeerreviewmaketitle

\section{Introduction}

\IEEEPARstart{T}{he} past few years have witnessed the rapid development of 5G communication technologies, \textcolor{black}{with the global deployment of over 3 million 5G base stations by 2022. However, the growing scale and intricacy of these networks pose challenges in terms of testing and deployment of new technologies. Directly integrating innovative technologies into the physical network can potentially disrupt normal operations, resulting in detrimental delays or even network failures, outcomes unacceptable to network operators. To tackle this issue, Digital Twin (DT) for 5G and beyond networks has been proposed and received increasing attention \cite{wu2021digital}. }

The concept of the DT was first formulated by Grieves and
Vickers \cite{grieves2017digital}, comprising a real space, a virtual space, and a data link \textcolor{black}{connecting the two spaces}. In recent years, DT has been used to simulate complex systems in fields such as aviation, manufacturing, and architectural design. \textcolor{black}{Similarly for 5G}, DT enables a software replica of the 5G physical network, which is instrumental in building flexible testbed facilities with high availability and accelerating the deployment of new technologies \cite{nguyenDigitalTwin5G2021}. 
Foreseeing the huge potential of DT, academia has identified it as one of the key enabling technologies for 5G and beyond wireless communication networks 
\cite{you6GWirelessCommunication2021, khanDigitalTwinEnabled6GVision2022}, \textcolor{black}{and the industry is proactively conducting research in this area, with participation from leading telecommunication companies such as Huawei \cite{huawei2020} and Ericsson \cite{ohlen2022}, as well as internet service providers like Bell Canada, China Mobile, UScellular, and Vodafone \cite{ngmn2022}.}

Most of the existing 5G network DTs are built for network optimization and resource scheduling \textcolor{black}{\cite{dong2019deep,wang2020graph,lu2020communication,liao2022cloud,zhou2021secure}}. \textcolor{black}{However, there are only limited DTs modeling 5G core network functions, such as \cite{mozoB5GEMINIAIDrivenNetwork2022} and \cite{vakaruk2021digital}, with 5G core implementations based directly on open-source projects. Although these projects can realize most 5G core features, significant differences remain compared to physical networks in terms of protocol versions, configurations, etc., necessitating extensive additional development. Furthermore, the complexities of heterogeneous physical networks across scenarios render faithfully replicating the physical network via programming time-consuming and labor-intensive.}

\textcolor{black}{Motivated by the limitations of existing DT modeling methods, this paper seeks an alternative approach that is compatible with different physical networks and avoids individual programming for each unique scenario. The current 5G core implements the control and user plane separation (CUPS), where the user plane is solely responsible for network user traffic. Therefore, the main challenge for modeling lies in the control plane, which integrates the vast majority of network functions. Consequently, this paper specifically focuses on the modeling of the control plane.}

\textcolor{black}{In recent years, deep learning (DL) methods have demonstrated remarkable success in various domains, particularly in computer vision (CV) and natural language processing (NLP). While classic CV models and solutions have found extensive applications in the communications field \cite{wen2018deep, ye2017power}, the utilization of NLP technologies remains limited due to disparities in data formats and application scenarios. Drawing inspiration from dialogue systems in NLP, we attempt to recognize the signaling interactions within the 5G core as dialogues between two planes. By leveraging DL models and training them on captured signaling data from specific interfaces, we aim to model the control plane functions through a data-driven methodology.}

\subsection{Related Work}
\subsubsection{5G network digital twin}
Although the concept of the DT has been proposed for a long time, the 5G network DTs are still in their nascent stage\textcolor{black}{, and only a few 5G network DT systems have been implemented so far. \citet{mozoB5GEMINIAIDrivenNetwork2022} proposed B5GEMINI, comprising primarily a virtual 5G core incorporating multiple network functions deployed in a distributed manner through virtual machines, a traffic generation module emulating the user equipment behavior, a bidirectional pipeline enabling real-time synchronization between the real and virtual networks, and an AI module for optimization and prediction. The implementation of network functions in the virtual 5G core is based on the Free5GC project. Similarly, with the purpose of training cyber security experts, \citet{vakaruk2021digital} presented a DT platform, SPIDER cyber range, based on Free5GC.}
\subsubsection{5G core simulation}
With the development of software-defined networking (SDN) and network functions virtualization (NFV), 5G networks no longer rely on monolithic components, thus open-source platforms for 5G core are receiving increasing attention. \textcolor{black}{One notable project is Free5GC, initiated by a research team primarily from the National Chiao Tung University \cite{free5gc}. They migrated the 4G Evolved Packet Core (EPC) into the 5G core Service-Based Architecture (SBA) in January 2019 and implemented a fully operational 5G core conforming to the 3rd Generation Partnership Project (3GPP) Release 15 (R15) in April 2020. Similarly, \citet{open5gs} led the team to establish the Open5GS, a C-language implementation of 5G core based on Release 16. The industry has also introduced specialized equipment like Spirent's Landslide for 5G core simulation. However, both open-source platforms and dedicated devices require additional development or extensive configuration to replicate a physical network. According to the information we have, there is a lack of solutions that can automate the modeling or configuration.}

\subsubsection{DL-based dialogue system}
\textcolor{black}{Dialogue system is a widely studied NLP task with promising real-life applications, and most state-of-the-art frameworks are based on DL due to their outstanding performance. Although classic recurrent neural networks (e.g., Jordan-type RNN \cite{jordan1997}, Elman-type RNNs \cite{elman1990}, LSTM \cite{hochreiter1997}) possess the capacity to handle sequential data, they are insufficient as standalone dialogue systems due to their inherent limitation of rigidly conforming to a one-to-one input-output mapping. To address this, \citet{sutskever2014} proposed the sequence-to-sequence (Seq2Seq) model, using an encoder to map the entire input sequence into an intermediate vector and a decoder to further generate the output based on the vector, thereby accommodating variable source and target sequences lengths. \citet{bahdanau2016} introduced the attention mechanism to the Seq2Seq model, allowing the decoder to consider the relationships with each part of the encoded source sentence.} \citet{vaswani2017} proposed Transformer, which completely adopts attention mechanisms without any RNNs to achieve both local and global dependencies and more parallelization. The advent of the Transformer makes it feasible to train large pre-trained models in the NLP domain. \citet{devlinBERTPretrainingDeep2019} proposed BERT based on a bidirectional Transformer encoder, and \citet{radford} proposed GPT based on a unidirectional Transformer decoder, both of them possess the capability to adapt to new tasks after pretraining. In the field of chatbots, the latest version of GPT, ChatGPT, has demonstrated remarkable performance through a large-scale model with the assistance of human feedback.

\textcolor{black}{In the field of communications, dialogue-based approaches have already been explored for the core of cellular networks \cite{Pereira}, \cite{Pereira2}. However, these studies exclusively focused on Session Initiation Protocol (SIP) signaling messages, and only predicted the signaling type rather than the complete signaling data. As such, the models proposed in these papers are not applicable to the control plane modeling.}

\subsection{Contributions and Organization}
\textcolor{black}{In this paper, we propose two deep learning models, namely 5G-Seq2Seq and 5G-former, which aim to replicate the primary functionalities of the 5G core network using signaling data.} To the best of our knowledge, this is the first study of modeling the 5G core function in DT through a deep learning method rather than programming. \textcolor{black}{By leveraging these models, we are able to automatically and accurately replicate the core network control plane's major functions.} The main contributions of this work can be listed as follows.

\begin{itemize}
\item{We propose two different 5G control plane architectures for modeling its response behavior when receiving uplink signaling messages. \textcolor{black}{Instead of relying on traditional programming methods, these architectures enable data-driven modeling by utilizing captured signaling data from the interfaces between the control and user planes.}
}

\item{We present a solution that allows signaling messages with different amounts of information and in different interfaces to be interconverted with uniform \textcolor{black}{length-limited} vectors, making it feasible to construct signaling datasets for neural network training.}

\item{We deploy two DL models based on the Vanilla Seq2Seq model and the Transformer decoder respectively, \textcolor{black}{with structural modifications tailored for signaling data, which yield high prediction accuracy across diverse 5G procedures.}}
\end{itemize}

The rest of the paper is organized as follows. The system model and the proposed architectures are presented in Section II. In Section III, we introduce some preliminaries to our proposed method and elaborate on the solution to construct the dataset on signaling data. In Section IV, we introduce two DL-based models, \textcolor{black}{followed by the presentation of experimental results and performance evaluation in Section V. Finally, we discuss the conclusions and future directions in Section VI.}

\section{System Model} \label{System Model}
\begin{figure}[!t]
\centering
\includegraphics[width=0.4\textwidth]{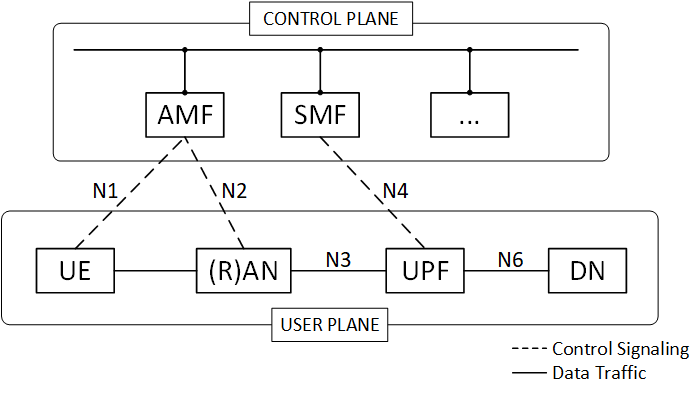}
\caption{Service-Based 5G core Network}
\label{5gc}
\end{figure}

\begin{figure}[!t]
\centering
\includegraphics[width=0.49\textwidth]{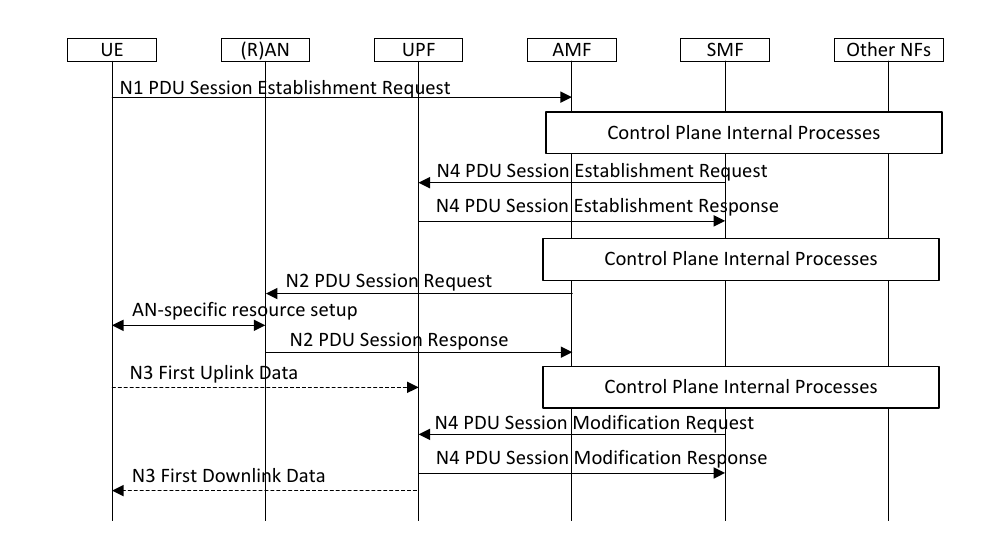}
\caption{UE requested PDU Session Establishment}
\label{pdu}
\end{figure}

We consider a 5G core network with control and user plane separation and a service-based architecture. As illustrated in Fig. \ref{5gc}, the main network functions of the control plane are the access and mobility management function (AMF) and the session management function (SMF), which are responsible for access control and session management respectively. The control plane also includes a number of network functions that assist AMF and SMF, including authentication server function (AUSF), unified data management (UDM), etc. In terms of the user plane, the user plane function (UPF) is deployed to process and forward data traffic from users, and the N3 and N6 interfaces are used to link the radio access network (RAN), UPF, and data network (DN). Through N1, N2, and N3 interfaces, the control plane achieves control of the user plane and user equipment(UE) by sending and receiving control signaling. Fig. \ref{pdu} shows an example of a registered UE establishing a PDU session via interfaces including N1, N2, N3, and N4.

\textcolor{black}{By drawing an analogy to a dialogue system, we can conceptualize the control plane and the user plane as participants engaged in a conversation. The signaling messages transmitted through the three interfaces (N1, N2, N4) can be seen as sentences within this dialogue. Consequently, we define the signaling from the user plane to the control plane as uplink signaling, denoted by $s_u$, while the signaling from the control plane to the user plane is referred to as downlink signaling, denoted by $s_d$. As the number of response signaling messages from the control plane may vary in response to the specific $s_u$, we employ $\mathbf{s_d}$ to represent the entirety of response messages. For the $i^{th}$ uplink signaling message $s_u^{i}$, the corresponding response can be expressed as follows:}
\begin{equation}
    \mathbf{s_d^{i}}=[\begin{array}{llll}s_d^{i,1} & s_d^{i,2} & \ldots & s_d^{i,j}\end{array}]^T,
\end{equation}
where $j$ denotes the number of response signaling messages. 

\begin{figure}[!t]
\centering
\includegraphics[width=0.25\textwidth]{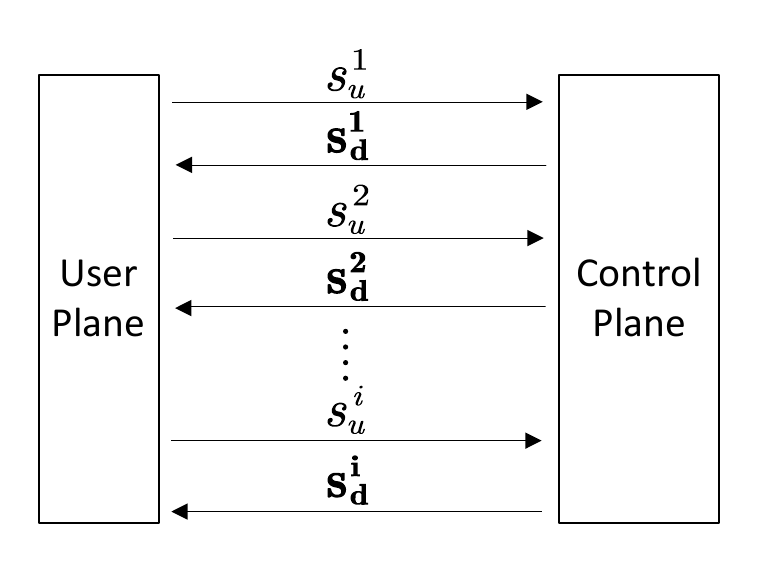}
\caption{Interaction between control plane and user plane in 5G core}
\label{sigmodel}
\end{figure}

In this way, we can simplify the core network structure into a dialogue system between the control and user plane, as shown in Fig. \ref{sigmodel}. The response $\mathbf{s_d^{i}}$ of the control plane when receiving the $i^{th}$ uplink signaling message $s_u^{i}$ can be modeled as
\begin{equation}
    \mathbf{s_d^{i}}=f(s_u^{i}, h^{i}),
    \label{sd}
\end{equation}
where $h^{i}$ represents the UE-related hidden state of the control plane at the $i^{th}$ interaction and $f(\cdot)$ is a nonlinear function that models the internal processes of the control plane. After the $i^{th}$ interaction, the hidden state $h^{i}$ will be updated to $h^{i+1}$ as
\begin{equation}
    h^{i+1}=g(s_u^{i},\mathbf{s_d^{i}}, h^{i}),
    \label{h}
\end{equation}
where $g(\cdot)$ is a nonlinear function that models the update of the hidden state. Furthermore, $h^{i}$ in \eqref{sd} can be replaced recursively by using \eqref{h}, thus \eqref{sd} can be rewritten as
\begin{equation}
    \begin{split}
     \mathbf{s_d^{i}}&=f(s_u^{i}, g(s_u^{i-1},\mathbf{s_d^{i-1}}, g(s_u^{i-2},\mathbf{s_d^{i-2}}, \ldots)))   \\
     & = z(s_u^{i}, s_u^{i-1}, \mathbf{s_d^{i-1}}, \cdots, s_u^{1},\mathbf{s_d^{1}},h^{1}),
    \end{split}
    \label{sd2}
\end{equation}
where $z(\cdot)$ is another nonlinear function, and $h^{1}$ denotes the initial state of UE before registration, which is a constant value.

\begin{figure}[!t]
\centering
\includegraphics[width=0.4\textwidth]{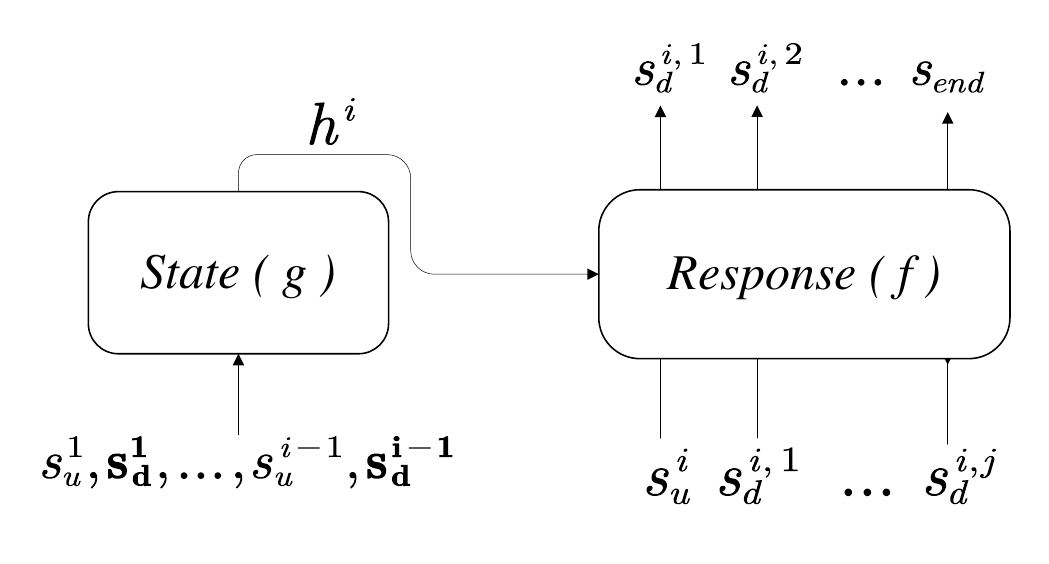}
\caption{State-based architecture}
\label{model1}
\end{figure}

\begin{figure}[!t]
\centering
\includegraphics[width=0.4\textwidth]{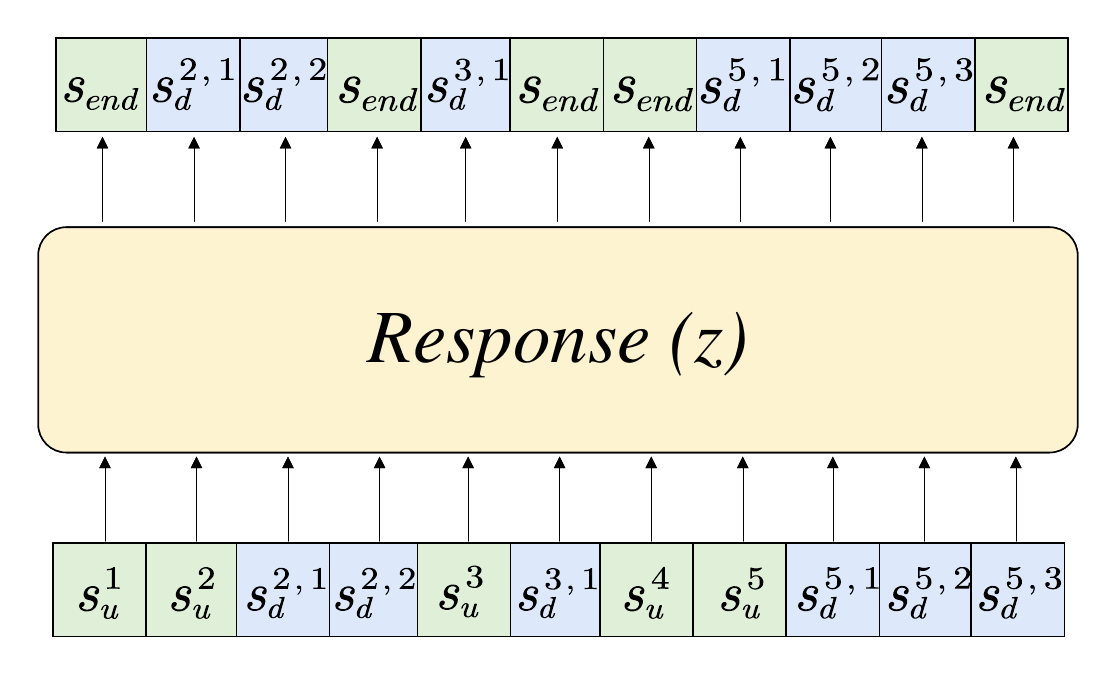}
\caption{Signaling-based architecture}
\label{model2}
\end{figure}

Drawing on the encoder-decoder structure in the Vanilla Seq2Seq model, we construct a state-based architecture for the control plane using \eqref{sd} and \eqref{h}. As shown in Fig. \ref{model1}, \textcolor{black}{the state block receives signaling input after each interaction to continuously update the hidden state $h^{i}$.} For the response block, both uplink signaling message $s_u^{i}$ and hidden state $h^{i}$ are utilized to predict the response $\mathbf{s_d^{i}}$. Since the length of $\mathbf{s_d^{i}}$ is not determined, an autoregressive approach is used to perform sequential prediction until the response block outputs the end of signaling messages $s_{end}$, which indicates that no more response is required.

\textcolor{black}{However, the Seq2Seq model, which utilizes RNN as its fundamental components, is unable to achieve a high level of parallelism, thereby limiting its potential for constructing large-scale models. Additionally, the utilization of the intermediate vector $h$ for state transfer may result in information loss, thus adversely affecting prediction accuracy.
Hence, we also propose an entirely signaling-based architecture using \eqref{sd2}, inspired by the GPT model. As shown in Fig. \ref{model1}, instead} of creating a constantly updated hidden state, this architecture responds directly by analyzing uplink signaling message $s_u^{i}$ and previous signaling messages through a single response block. Specifically, when $s_u^{i}$ is received, it will be fed into the block along with the past signaling for prediction. \textcolor{black}{This architecture also adopts an autoregressive approach, where the output is appended to the end of the sequence for subsequent prediction iterations until the block outputs $s_{end}$.}

\begin{figure*}[!t]
\centering
\includegraphics[width=1\textwidth]{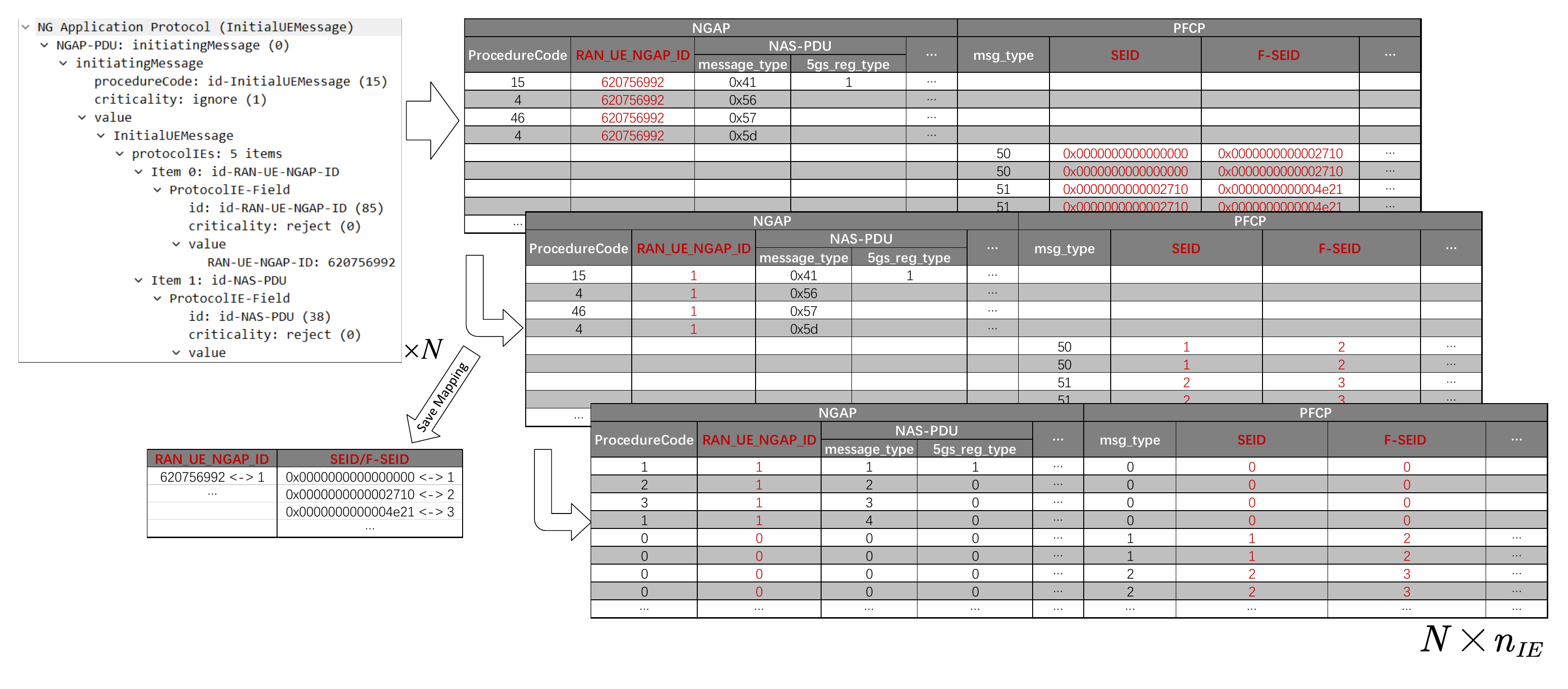}
\caption{\textcolor{black}{The proposed solution for data conversion}}
\label{transform}
\end{figure*}

\section{Preliminaries}
\textcolor{black}{\subsection{Signaling Capturing}}\label{capture}
In order to obtain diverse data, we use the Spirent C50 network tester to realize different UE behaviors and generate various signaling procedures (e.g., registration, de-registration, authentication, handover, PDU session establishment, PDU session modification, PDU session release). Moreover, Poisson distribution is applied to determine the number and time interval of sessions for each user, \textcolor{black}{thereby enhancing the diversity of signaling data. Signaling data can be captured in PCAP format by consistently monitoring the N1, N2, and N4 interfaces. Importantly, since the monitoring is solely conducted at the interfaces without any alterations to the control plane or user plane function, both the signaling capturing and subsequent data-driven modeling will not interfere with the normal operation of both planes.}

\textcolor{black}{\subsection{Pre-processing}}\label{dictionary}

\textcolor{black}{The captured signaling messages must be transformed into vectors to comply with the input format of the neural network. Employing conventional NLP techniques, as described in the aforementioned signaling prediction studies \cite{Pereira, Pereira2}, requires numbering each type of signaling message, creating a dictionary, and subsequently applying one-hot encoding based on the assigned numbers. However, in this task, adopting such a methodology would yield vectors of nearly infinite length, rendering them unfeasible for both storage and training. Specifically, each signaling message consists of multiple information elements (IEs), as shown in Fig. \ref{transform} (left). We define the number of potential types of IEs as $n_{IE}$, and the number of possible values for the $i^{th}$ IE as $n_{class}^{i}$, then the number of potential types of signaling messages, which also corresponds to the length of the one-hot encoded vector, is determined by $\prod_{i=1}^{n_{IE}}{{n^i}_{class}}$. Even supposing each IE has only two possible values, presence and absence, the length would still be $2^{n_{IE}}$. Not to mention the presence of IEs like \textit{RAN\_UE\_NGAP\_ID}, which assumes a 9-digit decimal value, coupled with the possibility of absence, engenders ${n^i}_{class}=10^9+1$ possible values for this single IE. This poses considerable challenges for further processing.}

\begin{figure}[t]
\centering
\includegraphics[width=0.46\textwidth]{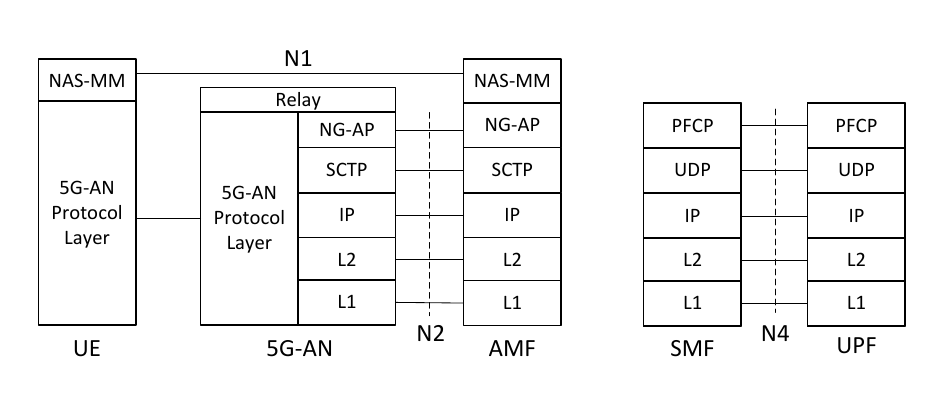}
\caption{Protocol stack in 5G core}
\label{stack}
\end{figure}

\textcolor{black}{Therefore, we propose a restorable transformation solution to reduce the length of the encoded vector as much as possible. Firstly, we use Wireshark software to decode the signaling messages. Fig. \ref{stack} shows Protocol Stacks on three interfaces. Since each protocol layer has strict rules for encapsulation and decapsulation, which is not the focus of this paper, we only keep payload data. Subsequently, we transformed the payload data, which is organized in a tree data structure, into a table with a total of $n_{IE}$ columns, with each column representing a distinct type of IE, as illustrated in Fig. \ref{transform} (right). Dictionaries could be created for each column, enabling the translation of signaling messages into natural number vectors and their reconstruction from vectors. In these dictionaries, `0' signifies the absence of the corresponding IE in the current signaling message, while `1, 2, 3...' respectively represent each possible value for the corresponding IE it is present. For example, the dictionary for the first column will be: \{`ProcedureCode:Null': 0, `ProcedureCode:15': 1, `ProcedureCode:4': 2, `ProcedureCode:46': 3 ...\}. The length of the $i^{th}$ dictionary (the number of key-value pairs in a dictionary), corresponds to the number of possible values for the $i^{th}$ IE, i.e. $n_{class}^{i}$. Then one-hot encoding can be applied to each column, and the resulting encoded vectors can be concatenated to obtain the final vector representing the entire signaling message. At this time, the vector length will be $\sum_{i=1}^{n_{IE}}{{n^i}_{class}}$, which is much more feasible for processing compared to $\prod_{i=1}^{n_{IE}}{{n^i}_{class}}$. }

\begin{algorithm*}[t]
	\caption{IE replacement in training set}\label{al1}
	\KwIn{unmodified signaling messages $\mathbf{s}$}
	\For{$i \leftarrow 1$ \KwTo $n_{IE_2}$}{
            $M^i = \left[m^{i,1},\dots,m^{i,n_{max}}\right] = \left[(I^{i,1},R^{i,1}),\dots,(I^{i,n_{max}},R^{i,n_{max}})\right]$\;\tcp{For $i^{th}$ type {\uppercase\expandafter{\romannumeral 2}} IE, $I$ denotes the original value, $R$ denotes the relative value, and $M^i$ is a list used to preserve the mapping between two values}
            \For{$j \leftarrow 1$ \KwTo $n_{max}$}
            {$I^{i,j}= Null$\;
            $R^{i,j}= j$\;}
	}
	\For{$k \leftarrow 1$ \KwTo $n_{sig}$}
        {
            Find all type {\uppercase\expandafter{\romannumeral 2}} IEs in the $k^{th}$ signaling message $s^k$\;
            \For{$i \leftarrow 1$ \KwTo $n_{IE_2}$}
            {            
                \If{$IE_2^i \neq Null$}
                {
                \eIf{$IE_2^i$ \textbf{in} $M^i.I$}
                    { 
                    find index $j$ to make $I^{i,j} = IE_2^i$\;
                    $IE_2^i \leftarrow R^{i,j}$\; 
                    \tcp{replace $i^{th}$ unusable IE in message with relative value}
                    move $m^{i,j}$ to the end of the list $M^i$\;
                    \tcp{The more recent the IE used, the closer it is to the end of the list}
                    }
                    {
                    append $m=(IE_2^i,R^{i,1})$ to the end of the list $M^i$\;
                    $IE_2^i \leftarrow R^{i,1}$\;
                    remove $m^{i,1}$ from the list\;
                    \tcp{When the relative values are all occupied, the new IE will overwrite the least used IE}
                    }
                }
            }
	}
	\KwOut{$\mathbf{s}$} 
\end{algorithm*}

\textcolor{black}{Although the length of the encoded vector has been significantly reduced, it remains oversized due to the presence of certain IEs like \textit{RAN\_UE\_NGAP\_ID}, which requires further pre-processing. Generally, the IEs contained in the signaling can be roughly divided into two categories. The first category is IEs with a limited range of values, namely type {\uppercase\expandafter{\romannumeral 1}} IEs, such as \textit{ProcedureCode} and \textit{msg\_type}. The second category is IEs with a wide range of values, namely type {\uppercase\expandafter{\romannumeral 2}} IEs, marked red in Fig. \ref{transform},  such as ID-related IEs (e.g., \textit{RAN\_UE\_NGAP\_ID}, \textit{SEID}) and authentication-related IE, which varies constantly and cannot be utilized directly in training. Hence, we perform a mapping operation on these IEs to replace them with relative values according to the order in which they are first received by the control plane. As shown in Algorithm \ref{al1}, we construct several lists for each type {\uppercase\expandafter{\romannumeral 2}} IE to save the mappings between the original IE values and the relative ones. By iterating through each signaling message, the original values are replaced while the mapping lists are continuously updated. Besides, certain IEs, for instance, the illegal \textit{IMSI}, need to be verified by UDM, which is beyond the capacity of DL models. Thus, these IEs can be labeled during the replacement process. After this operation, the number of types for type {\uppercase\expandafter{\romannumeral 2}} IEs are all restricted to $n_{max}$, and the length of encoded vector is reduced to $\sum_{i=1}^{n_{IE}}{l_i}$, where
\begin{equation}
l_i = \begin{cases} {n^i}_{class} & \text { if the $i^{th}$ IE $\in$ type {\uppercase\expandafter{\romannumeral 1}} IEs } \\ 
n_{max} & \text { if the $i^{th}$ IE $\in$ type {\uppercase\expandafter{\romannumeral 2}} IEs } \end{cases}.
\end{equation}
The complete data conversion process from signaling messages to vectors is illustrated in Fig. \ref{transform}. }

\begin{figure}[t]
\centering
\includegraphics[width=0.49\textwidth]{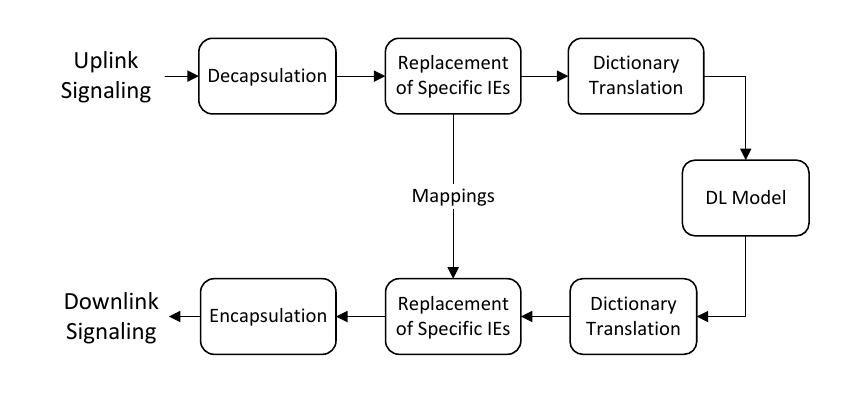}
\caption{\textcolor{black}{Signaling response via DL model}}
\label{response}
\end{figure}

Since each step of the conversion mentioned above is reversible, the downlink signaling messages can be obtained by stepwise restoration based on the prediction from the DL model, as depicted in Fig. \ref{response}. The detailed signaling response method via the DL model is shown in Algorithm \ref{al2}. The upper loop is used to substitute the type {\uppercase\expandafter{\romannumeral 2}} IEs in the messages before feeding them into the DL model, while the lower loop is used to continuously predict until the output is $S_{end}$, meanwhile backfilling the original values of the type {\uppercase\expandafter{\romannumeral 2}} IEs into the messages. The mapping lists are preserved and updated consistently in both loops so as to restore type {\uppercase\expandafter{\romannumeral 2}} IEs. If an output value is not in the mapping, the corresponding IE will be allocated (e.g. \textit{SEID} allocation) or calculated (e.g. hash-based message authentication) according to preset values and other IEs. 

\SetKwRepeat{Do}{do}{while}
\begin{algorithm*}[t]
	\caption{Signaling response via DL model}\label{al2}
	\KwIn{Uplink signaling message $s_u$, Mapping lists $M^1,\dots,M^{n_{IE_2}}$}
	unpack $s_u$ with the corresponding protocol stack\;
        \For{$i \leftarrow 1$ \KwTo $n_{IE_2}$}
        {        
            \If{$IE_2^i \neq Null$}
            {
            \eIf{$IE_2^i$ \textbf{in} $M^i.I$}
            { 
                find index $j$ to make $I^{i,j} = IE_2^i$\;
                $IE_2^i \leftarrow R^{i,j}$\; 
                move $m^{i,j}$ to the end of the list $M^i$\;
            }
            {
                append $m=(IE_2^i,R^{i,1})$ to the end of the list $M^i$\;
                $IE_2^i \leftarrow R^{i,1}$\;
                remove $m^{i,1}$ from the list\;
            } 
            }
        }
        convert $s_u$ into a vector with the dictionaries of each IE\;
        $s_{in} \leftarrow s_u$, $\mathbf{s_d} \leftarrow \left[\,\right]$\;
        \Do{$s_{out} \neq s_{end}$}{
        feed $s_{in}$ into the DL model to get the output $s_{out}$\;
        \If{$s_{out} \neq s_{end}$}
        {
            $s_{in} \leftarrow s_{out}$\;
            convert $s_{out}$ into IE values with the dictionaries of each IE\;
            \For{$i \leftarrow 1$ \KwTo $n_{IE_2}$}
            {\If(\tcp*[h]{Equivalent to ($IE_2^i$ \textbf{in} $M^i.R$)}){$IE_2^i \neq Null$}
            {
            \eIf{$IE_2^i \neq R^{i,1}$}
            { 
                find index $j$ to make $R^{i,j} = IE_2^i$\;
                $IE_2^i \leftarrow I^{i,j}$\; 
                move $m^{i,j}$ to the end of the list $M^i$\;
            }
            {
                allocate or calculate $IE_{new}$ according to preset values and other IEs\;
                append $m=(IE_{new},R^{i,1})$ to the end of the list $M^i$\;
                $IE_2^i \leftarrow IE_{new}$\;
                remove $m^{i,1}$ from the list\;
            } 
            }
            }
            encapsulate $s_u$ with the corresponding protocol stack\;
            append $s_{out}$ to the end of the list $\mathbf{s_d}$\;
        }}

	\KwOut{\textcolor{black}{Downlink signaling messages $\mathbf{s_d}$, Mapping lists $M^1,\dots,M^{n_{IE_2}}$}}
\end{algorithm*}

With this solution, each signaling message can be interconverted with $n_{IE}$-length vector $s \in \mathbb{N}^{n_{IE}}$ \textcolor{black}{(not encoded yet; after distributed one-hot encoding, the length of the vector will become $\sum_{i=1}^{n_{IE}}{l_i}$), which makes it practical to construct a dataset for training neural networks.}

\subsection{Dataset Construction and Splitting}\label{dataset}
To build datasets suitable for a specific model, signaling first needs to be divided into $s_u$ and $s_d$ according to their respective source and destination addresses. Subsequently, the components $s_u^{i}$ and $\mathbf{s_d^{i}}$ are aligned based on temporal, identification, and status-related information.

\subsubsection{Dataset for state-based model} The dataset for such a model consists of encoder inputs, decoder inputs, and decoder outputs.  \textcolor{black}{For an uplink signaling vector $s_u^{i}$, taking Fig. \ref{model1} as an example, the past signaling vectors ranging from $(s_u^{1}, \mathbf{s_d^{1}})$ to $(s_u^{i-1} ,\mathbf{s_d^{i-1}})$ are all stacked as the encoder input $i_{en}$. $s_u^{i}$ and the current response $\mathbf{s_d^{i}}$ are concatenated as the decoder input $i_{de}$. Then $i_{de}$ can be shifted left and appended with $s_{end}$ to obtain the decoder output $o_{de}$.} Apart from this, zero-padding, a technique of inserting empty vectors $s_{pad}$ that are dismissed by the model, is used to unify the length of the dataset. Mathematically, they are defined as
\begin{equation}
    i_{en} = [\begin{array}{llllllll}
    s_u^{1} &\mathbf{s_d^{1}} &\ldots & s_u^{i-1} &\mathbf{s_d^{i-1}} & s_{pad}&\ldots & s_{pad}
    \end{array}]^T,
\end{equation}
\begin{align}
    i_{de} &= [\begin{array}{lllll}s_u^{i}&\mathbf{s_d^{i}}&s_{pad}&\ldots&s_{pad}
    \end{array}]^T,\\
    o_{de} &= [\begin{array}{lllll}
    \mathbf{s_d^{i}} &s_{end} & s_{pad}&\ldots& s_{pad}
    \end{array}]^T.
\end{align}
The length of $i_{en}$ is fixed to $n_{all}$ based on the number of all messages before de-registration, while the length of $i_{de}$ and $o_{de}$ is fixed to $n_{res}$, which is relevant to the maximum number of messages in a single response. This operation is performed for every uplink message to construct dataset, including encoder inputs $I_{en} \in \mathbb{N}^{n_{ul} \times n_{all} \times n_{IE}}$, decoder inputs $I_{de} \in \mathbb{N}^{n_{ul} \times n_{res} \times n_{IE}}$, and decoder outputs $O_{de} \in \mathbb{N}^{n_{ul} \times n_{res} \times n_{IE}}$, where $n_{ul}$ denotes the number of captured uplink signaling messages.

\subsubsection{Dataset for signaling-based model} \textcolor{black}{Compared to the aforementioned approach, constructing a dataset for the signaling-based model is relatively straightforward.} Due to the unidirectionality of the generative model, each signaling vector can only be analyzed with the left-side past signaling vectors. Therefore, instead of individually analyzing each response, we can directly stack all the signaling vectors before de-registration to form the input. \textcolor{black}{To obtain the output, we perform a left-shift operation on the input sequence and replace all $\mathbf{s_u^{i}}$ components with $S_{end}$.} The zero-padding technique is still required in this dataset. The input and output can be represented as
\begin{align}
    i &= [\begin{array}{llllllll}
    s_u^{1} &\mathbf{s_d^{1}} &\ldots & s_u^{last} &\mathbf{s_d^{last}} & s_{pad}&\ldots & s_{pad}
    \end{array}]^T,\\
    o &= [\begin{array}{llllllll}
    \mathbf{s_d^{1}} &s_{end} &\ldots &\mathbf{s_d^{last}} &s_{end}& s_{pad}&\ldots & s_{pad}
    \end{array}]^T.
\end{align}
This operation is performed for each complete interaction process from registration to deregistration to build the dataset, including inputs $I \in \mathbb{N}^{n_{proc} \times n_{all} \times n_{IE}}$, and outputs $O \in \mathbb{N}^{n_{ul} \times n_{all} \times n_{IE}}$, where $n_{proc}$ denotes the number of complete processes.

Apart from the IEs, the time interval between each message and registration message is also saved in the same format. The dataset is partitioned into training, validation, and test sets at a ratio of 3:1:1. Due to the sequential feature of the data, \textcolor{black}{shuffling within each set of signaling messages between registration and deregistration should be avoided.}

\section{Realization of Deep Learning Models}
Two deep learning models, \textcolor{black}{namely} 5GC-Seq2Seq and 5GC-former, are proposed based on the architectures proposed in Section \ref{System Model}.

\subsection{5GC-Seq2Seq}
\begin{figure}[!t]
\centering
\includegraphics[width=0.4\textwidth]{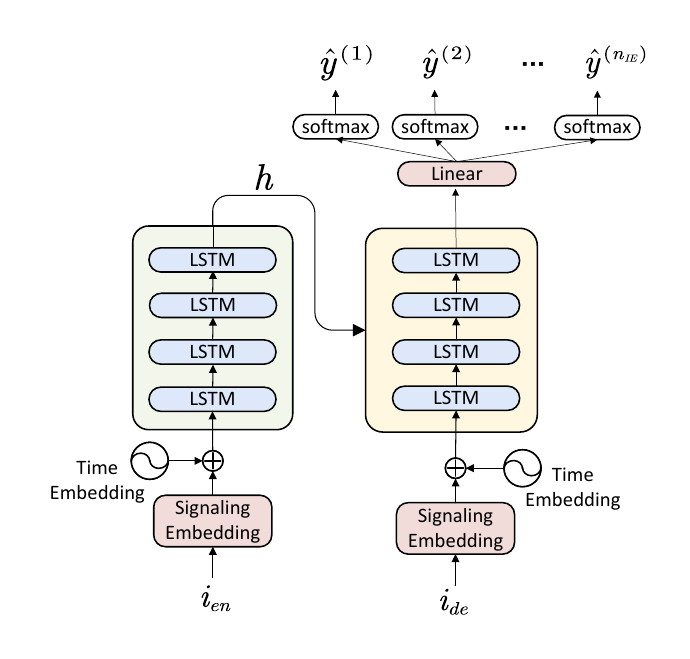}
\caption{5GC-Seq2Seq Model}
\label{seq2seq2}
\end{figure}

\begin{figure}[!t]
\centering
\includegraphics[width=0.49\textwidth]{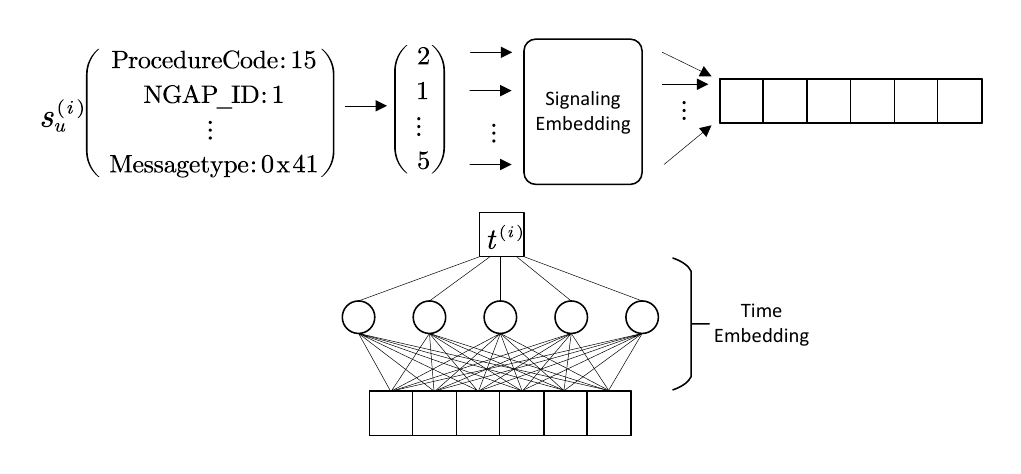}
\caption{Embedding}
\label{embed}
\end{figure}

Fig. \ref{seq2seq2} illustrates the structure of the 5GC-Seq2Seq model, which has the following three fundamental building blocks.
\subsubsection{Embedding Layer}
This layer converts the data into vectors of dimension $d_{model}$, which is the same dimension as the encoder/decoder. In NLP, this operation is known as Embedding, which uses a trainable embedding layer to convert words to vectors (Word2Vec) based on the corresponding dictionary number. \textcolor{black}{However, in contrast to words, signaling messages contain multiple IEs, with each IE having its own associated dictionary. Therefore, the IEs in the signaling need to be embedded separately and then summed up to create vectors of dimension $d_{model}$, as shown in Fig. \ref{embed}. Unlike adjacent words with fixed intervals, signaling messages have different time intervals, so the time difference between each message and the initial registration message also requires embedding. The embedding operation is performed by a feedforward neural network.}

\begin{figure}[!t]
\centering
\includegraphics[width=0.3\textwidth]{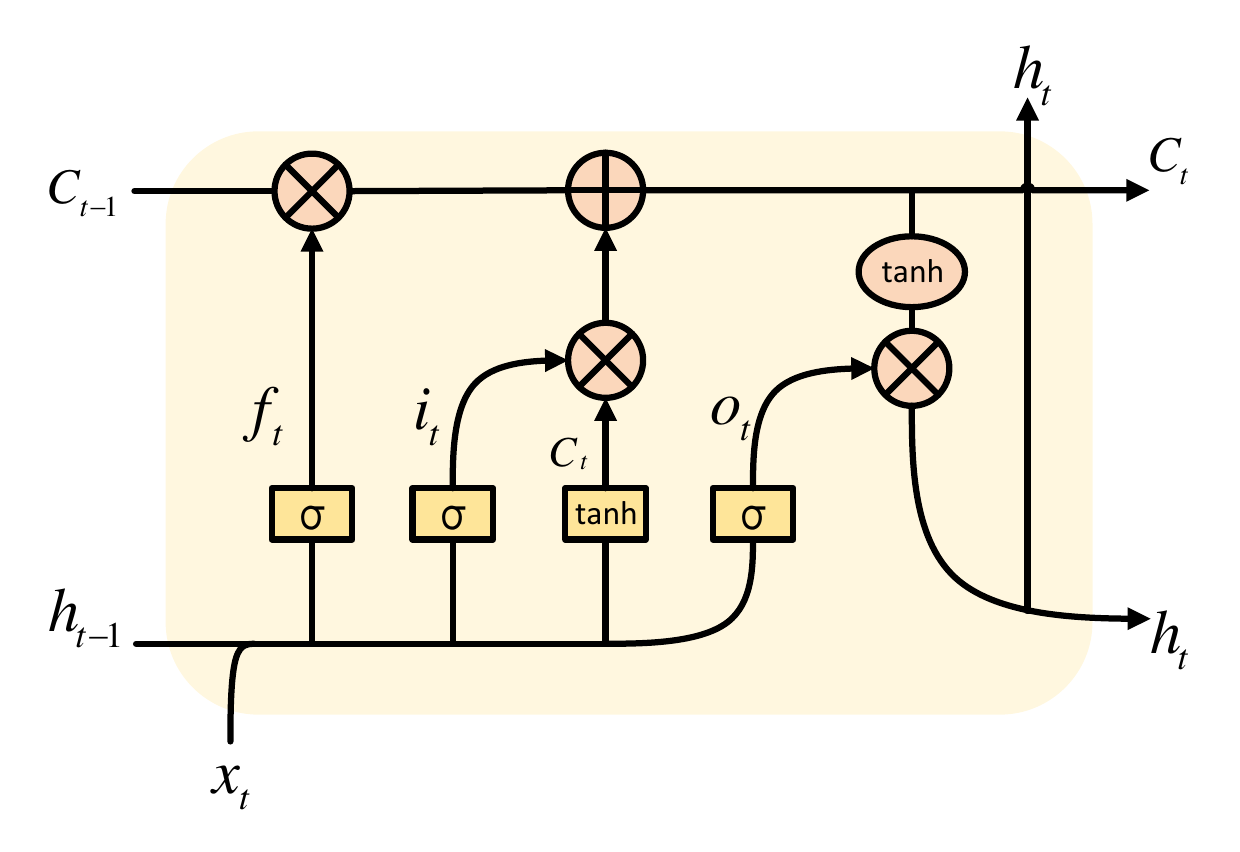}
\caption{LSTM cell}
\label{lstm}
\end{figure}

\subsubsection{LSTM encoder-decoder}
\textcolor{black}{Both the encoder and decoder in this model are composed of a stack of $N$ identical LSTM layers.} The structure of a single LSTM cell at time step $t$ is depicted in Fig. \ref{lstm}, along with the following equations:
\begin{equation}
    \begin{cases}
\mathbf{i_t} & =\sigma\left(\mathbf{w_i} \cdot\left[\mathbf{h_{t-1}}, \mathbf{x_t}\right]+\mathbf{b_i}\right), \\
\mathbf{o_t} & =\sigma\left(\mathbf{w_o} \cdot\left[\mathbf{h_{t-1}}, \mathbf{x_t}\right]+\mathbf{b_o}\right), \\
\mathbf{f_t} &=\sigma\left(\mathbf{w_f} \cdot\left[\mathbf{h_{t-1}}, \mathbf{x_t}\right]+\mathbf{b_f}\right),\\
\tilde{\mathbf{C}}_\mathbf{t} & =\tanh \left(\mathbf{w_C} \cdot\left[\mathbf{h_{t-1}}, \mathbf{x_t}\right]+\mathbf{b_C}\right),\\
\mathbf{C_t} & =\mathbf{f_t} * \mathbf{C_{t-1}}+\mathbf{i_t} * \tilde{\mathbf{C}}_\mathbf{t} ,\\
\mathbf{h_t} & =\mathbf{o_t} * \tanh \left(\mathbf{C_t}\right),
\end{cases}
\end{equation}
where $\mathbf{i_t}, \mathbf{o_t}, \mathbf{f_t}, \tilde{\mathbf{C}}_\mathbf{t}, \mathbf{C_t}, \mathbf{h_t}$ are the input gate, output gate, forget gate, cell candidate, cell state, and cell output respectively; $\sigma(\cdot)$ is the sigmoid function; $*$ denotes element-wise product; $\mathbf{w_{i}},\mathbf{w_o},\mathbf{w_f},\mathbf{w_C}$ and $\mathbf{b_i},\mathbf{b_o},\mathbf{b_f},\mathbf{b_C}$ are trainable weights and biases of corresponding gates. The input gate $\mathbf{i_t}$ determines the contribution of input $\mathbf{x_t}$ in updating the cell state $\mathbf{C_t}$, while the forget gate $\mathbf{f_t}$ controls the recession of the previous state. When the state has been updated, the output of the cell can be calculated through the output gate $\mathbf{o_t}$. 

\subsubsection{Output Layer}
The output layer consists of linear projections and softmax functions, which generate the probability vectors of different IEs as
\begin{equation}
    P^{i} = \operatorname{softmax}(O W^{i}),
\end{equation}
where $P^{i}$ is the probability vector of the $i^{th}$ IE, $O$ is the output of decoder, the projections are weight matrices $W^{i} \in \mathbb{R}^{d_{model}\times n^{i}_{class}}$, $n_{class}^{i}$ is the number of classes in the $i^{th}$ IE. The greedy algorithm is used to predict the values of IEs as follows.
\begin{equation}
    \hat{y}^{i} = \arg \max(P^{i}).
\end{equation}

\subsection{5GC-former}
\begin{figure}[!t]
\centering
\includegraphics[width=0.49\textwidth]{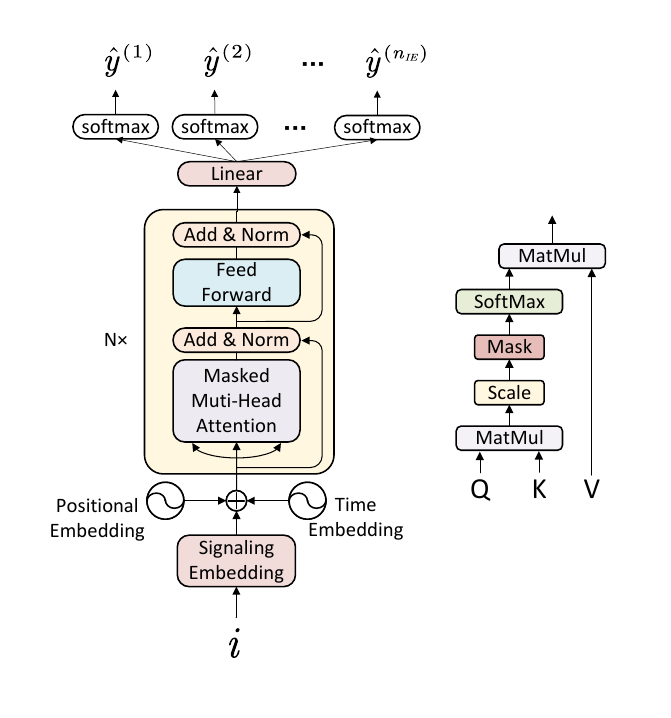}
\caption{(Left) 5GC-former Model. (Right) Attention cell.}
\label{Transformer}
\end{figure}

Fig. \ref{Transformer} depicts the structure of the 5GC-former model, which consists of the following three fundamental building blocks.


\subsubsection{Embedding Layer} \textcolor{black}{Due to the analogous input structure, the embedding layer of 5GC-former bears resemblance to that of 5GC-Seq2Seq in both the signaling embedding and time embedding. Nonetheless, the attention-based model forgoes the utilization of RNNs and instead processes all input data in parallel, thereby sacrificing the inherent ability of LSTMs to capture temporal dependencies. To enable the model to exploit the sequential order of the input sequence, the position of the input vector in the sequence also needs to be embedded.}

\subsubsection{Attention decoder} In this layer, we use masked self-attention, an intra-attention mechanism that relates different positions of a sequence and compute representations \cite{vaswani2017}, to explore the temporal dynamics of signaling messages. \textcolor{black}{First}, the embedded input matrix $I$ is transformed into query matrix $Q$, key matrix $K$, and value matrix $V$ using linear projections as
\begin{equation}
    \begin{cases}
Q&=I W_q \\
K&=I W_k \\
V&=I W_v
\end{cases},
\end{equation}
where the projections are weight matrices $W_q \in \mathbb{R}^{d_{model}\times d_q}$, $W_k \in \mathbb{R}^{d_{model}\times d_k}$, $W_v \in \mathbb{R}^{d_{model}\times d_v}$. Then query $Q$ is dot-product with key $K$ to calculate the relationships between vectors, and the result needs to be normalized by the square root of the hidden dimension as well as the softmax function. Due to the unidirectionality of the generative model, attention needs to be masked here so that each vector can only be associated with the previous vectors. In the end, the \textcolor{black}{values are} weighted summed according to the relationships. The whole process is depicted in Fig. \ref{Transformer} (right), and defined as
\begin{equation}
    \operatorname{Attention}(Q, K, V)=\operatorname{softmax}\left(\frac{Q K^T}{\sqrt{d_k}}\right) V.
\end{equation}

\textcolor{black}{Furthermore, the model employs multi-head attention, which executes multiple attentive functions concurrently and then \textcolor{black}{synthesizes} the outcomes. This enables the model to effectively focus on information from various representation subspaces and positions simultaneously. The concept of multi-head attention is defined as follows:}
\begin{equation}
\begin{aligned}
\operatorname{MultiHead}(Q, K, V) & =\operatorname{Concat}(\operatorname{head}_1, \ldots, \operatorname{head}_{\mathrm{h}}) W_o \\
\text { where \textcolor{black}{$\operatorname{head}_i$} } & =\operatorname{Attention}(Q W_q^{i}, K W_k^{i}, V W_v^{i})
\end{aligned}
\end{equation}
where the projections are weight matrices $W_q \in \mathbb{R}^{d_{model}\times d_q}$, $W_k \in \mathbb{R}^{d_{model}\times d_k}$, $W_o \in \mathbb{R}^{hd_v\times d_{model}}$, and $h$ is the number of heads.

Apart from the attention network, the decoder also adopts the feed-forward network to improve the fitting ability of this model. The feed-forward network consists of two linear transformations with a ReLU activation function and is defined as
\begin{equation}
    \operatorname{FFN}(x)=\max \left(0, x W_1+b_1\right) W_2+b_2,
\end{equation}
where the projections are weight matrices $W_1 \in \mathbb{R}^{d_{model}\times d_{ff}}$ and $W_2 \in \mathbb{R}^{d_{ff} \times d_{model}}$, $d_{ff}$ is the dimension of the inner-layer, and $x$ denotes the input of the feed forward network.

Both the feed-forward network and masked multi-headed attention network require residual connection \cite{he2016deep} as well as layer normalization \cite{ba2016layer} to prevent gradient explosion and vanishing when training deep networks.
\begin{equation}
    \text { SubLayerOutput }=\text{LayerNorm}(x+\text{SubLayer}(x))
\end{equation}

The decoder is composed of a stack of $N$ identical layers which consist of masked multi-head self-attention, feed-forward network, residual connection, and layer normalization.

\subsubsection{Output Layer}Due to the same format of the target vectors, the output layer of 5GC-former is the same as that of 5GC-Seq2Seq.

\subsection{Loss function}
In most of the classification tasks, the cross entropy loss function is used to measure the performance of the model, which is defined as
\begin{equation}
    \mathcal{L}_{\mathrm{CE}}=-\sum_{i=1}^n t_i \log \left(p_i\right),
\end{equation}
where $t_i$ is the truth label, $p_i$ is the softmax probability for the $i^{th}$ class, and $n$ is the number of classes. However, the output of models in this paper consists of multiple classification results, so the loss function is redefined as 
\begin{equation}
    \mathcal{L}=- \frac{1}{n_{IE}} \sum_{i=1}^{n_{IE}} \sum_{j=1}^{n_{class}^i} \mathbb{I}(j = y^i) \log \left(P^{i,j}\right).
    \label{loss}
\end{equation}
\textcolor{black}{Here, $y^i$ represents the $i^{th}$ value within the target vector, which can be translated into the value of the $i^{th}$ IE in the target message according to the corresponding dictionary, and $P^{i,j}$ represents the $j^{th}$ value in the probability vector of $i^{th}$ IE. $\mathbb{I}(\cdot)$ represents the indicator function, taking a value of 1 when its argument is true and 0 when false.}

\section{Implementation details and performance evaluation}
\subsection{Experimental Set-up}
\textcolor{black}{To get the dataset for the experiment, we first employ the signaling capturing methodology as described in Section \ref{capture} to get typical dialogues involving various quantities of PDU sessions and different procedures. Then, to augment the complexity of the signaling data for a more accurate simulation of real-world scenarios, we manipulate the time interval between different procedures to interleave them. For example, an additional PDU session request can be raised before the previously requested PDU session has been completely established. Meanwhile, the multi-user concurrency situation is also considered by enabling the interleaving of different users' signaling dialogues. A total of 90,766 signaling messages were obtained in the end, comprising 20.18\% registration signaling, 14.98\% deregistration signaling, 41.32\% PDU session signaling, 5.93\% handover signaling (intra-AMF), 11.34\% handover signaling (inter-AMF), and 5.24\% authentication signaling. Among the captured signaling messages, the number of potential types of IEs $n_{IE}$ is 676. We set the maximum types of relative values $n_{max}$ as 8, then employ the pre-processing method proposed in Section \ref{dictionary} to convert the signaling messages into vectors, and finally construct the dataset.}

The 5GC-Seq2Seq and 5GC-former designed in this paper are implemented based on \textcolor{black}{Python 3.9, Pytorch 1.10, CUDA 11.3,} and Numpy. The experimentation is performed on a commercial PC (i7-12700KF CPU, Windows 11 64-bit operating system, and 32 GB RAM) with a dedicated GPU (NVIDIA GeForce RTX 3080). 

\subsection{Performance Metrics}
The most common performance metrics in current machine learning tasks are accuracy, precision, recall, and F1-score. However, all these metrics are designed for classification tasks. \textcolor{black}{In the case of signaling messages, each IE has multiple possible values and each signaling message has multiple IEs, which is actually a multi-class multi-label classification task. Consequently, the evaluation metrics necessitate a redefinition to suit this particular scenario.}

Considering a single signaling message, the accuracy can be defined as
\begin{equation}
\text { Accuracy }=  \frac{\sum\limits_{i=1}^{n_{IE}} \mathbb{I}(y^{i} = \hat{y}^{i})}
{n_{IE}}.
\end{equation}
\textcolor{black}{Here, $y^{i}$ and $\hat{y}^{i}$ represent the $i^{th}$ values in the target and predictive vectors respectively, which can be translated into the values of the $i^{th}$ IE in the target and predicted message according to the corresponding dictionary. The metric can be further explained through an example: when the target message carries the IE \textit{ProcedureCode} with value 46, the first value $y^{1}$ in the target vector will be 3 according to the dictionary \{`ProcedureCode:46': 3\}. For a correct prediction $\mathbb{I}(y^{i} = \hat{y}^{i})$, the predicted value $\hat{y}^{1}$ must be 3. On the other hand, when the target message does not contain this IE, the first value $y^{1}$ will be 0 as per the dictionary \{`ProcedureCode:Null': 0\}, and the predictive value $\hat{y}^{1}$ must be 0 for a correct prediction.}

\textcolor{black}{However, due to the sparse distribution of the IEs in different signaling, the number of IEs absent far exceeds the genuinely carried IEs in the message, making the accuracy unsuitable for model evaluation in comparison to recall or precision. For the classification task, recall and precision are computed by dividing the categories into positive and negative samples. In terms of signaling messages, we define the value 0, which indicates that the IE is absent in the message, as negative samples, and all other values are considered positive samples. So in this task, recall represents the probability that the IEs carried in the target message are accurately predicted, and precision denotes the probability that the IEs in the predicted message are exactly carried in the target message.} The two metrics are defined as
\begin{align}
    \text { Recall }&=  \frac{\sum\limits_{i=1}^{n_{IE}} \mathbb{I}(y^{i} = \hat{y}^{i} \wedge y^{i} \neq 0) }
                {\sum\limits_{i=1}^{n_{IE}}  \mathbb{I}(y^{i} \neq 0)}\\
\text { Precision }&=  \frac{\sum\limits_{i=1}^{n_{IE}} \mathbb{I}(y^{i} = \hat{y}^{i} \wedge y^{i} \neq 0)}
                {\sum\limits_{i=1}^{n_{IE}}  \mathbb{I}(\hat{y}^{i} \neq 0)}.
\end{align}
F1-score is also introduced to take both precision and recall into account, which is defined as follows:
\begin{align}
     \text {F1-Score}=2 \cdot \frac{\text {Precision} \cdot \text {Recall}}{\text {Precision}+\text {Recall}}.
\end{align}
These metrics are calculated respectively for all valid signaling messages, i.e., non-padding messages, and then averaged to determine the overall metrics.

\subsection{Training}
In our experiment, we set the model dimension $d_{model} = 200$ and the number of sublayers $N = 6$ for both models, and set the number of heads $h = 4$ in multi-head attention as well as the dimensionality of inner-layer $d_{ff} = 400$ in the feed-forward network for 5GC-former in particular. The models are trained on the training set and validation set mentioned in Section \ref{dataset}, by using Adam optimizer \cite{kingma2014adam} with $\beta_1=0.9, \beta_2=0.99 \text { and } \epsilon=10^{-8}$. Due to the overfitting that occurred after 
hundreds of epochs in the experiment, dropout regularization, and early stopping are employed to increase training efficiency. 

\subsubsection{Dropout regularization}
Dropout \cite{srivastava2014dropout} is a regularization practice that randomly disregards certain nodes in layers during training, which can prevent overfitting by ensuring that there are no interdependencies between units. We apply dropout in both models and set the dropout rate as $0.1$.


\subsubsection{Early stopping}
\textcolor{black}{Rather than using a fixed number of training epochs, this study employs a dynamic approach where the training process can be terminated based on the validation loss. If the validation loss shows no improvement over several consecutive epochs, the training is halted prematurely. The model with the lowest validation loss is then saved, effectively preventing further overfitting. The early stopping patience for both models is set as $20$ (epochs).}

\begin{table}[]
    \centering
    \caption{Training parameters}
    \begin{tabular}{cc}
        \toprule \textbf{Parameter} & \textbf{Value}\\
        \midrule 
        Dimension of models & 200 \\
        Number of sublayers  & 6 \\
        Heads in multi-head attention & 4 \\
        Dimension of FFN & 400 \\
        Adam optimizer parameters & $(0.9,0.99,1e-8)$\\
        Batchsize & 8 \\
        Dropout rate & 0.1 \\
        Learning rate & 0.0001 \\
        Early stopping patience & 20 \\
        \bottomrule
    \end{tabular}
    \label{tab0}
\end{table}

\begin{figure}[!t]
\centering
\includegraphics[width=0.4\textwidth]{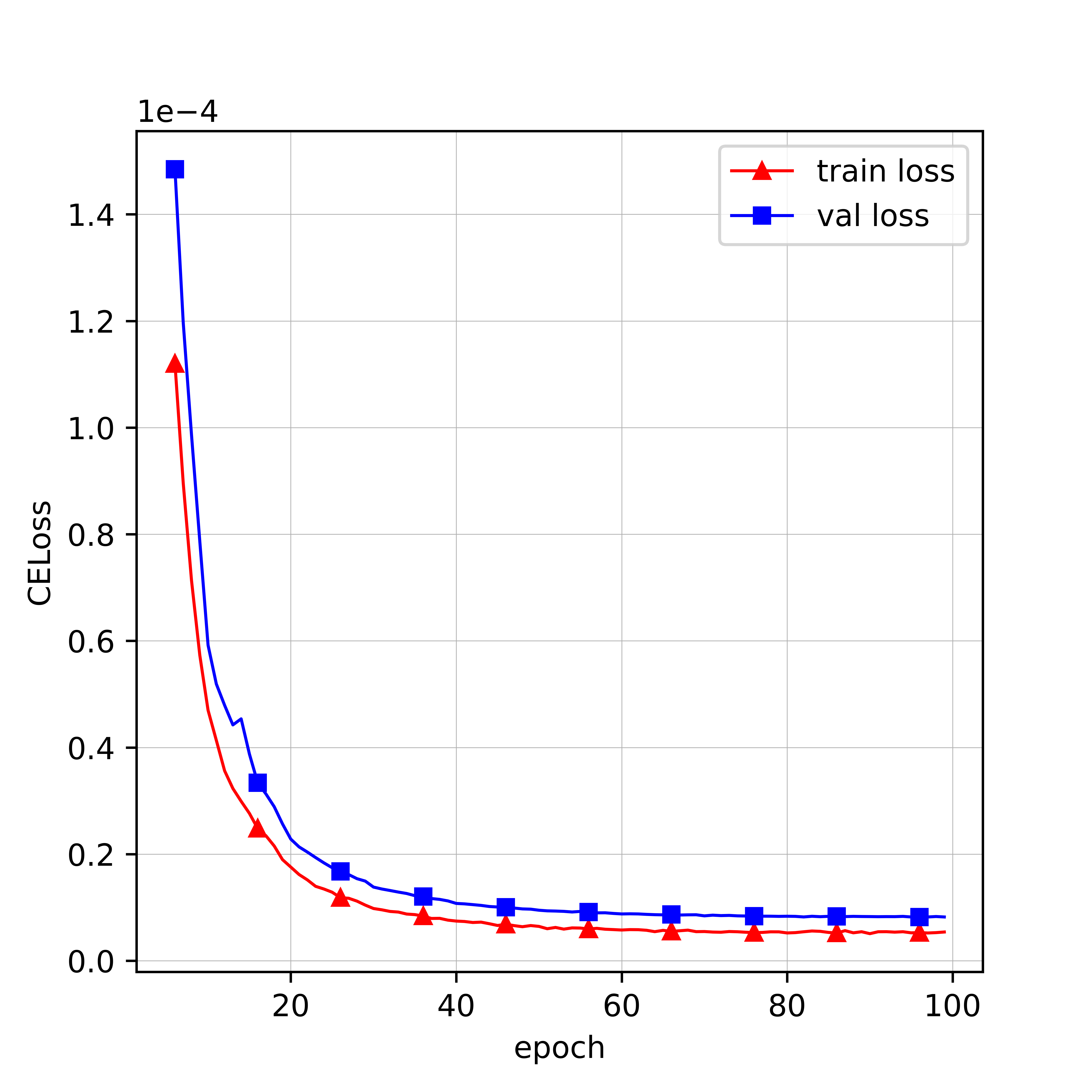}
\caption{The loss curve of the 5GC-Seq2Seq model}
\label{losss}
\end{figure}

\begin{figure}[!t]
\centering
\includegraphics[width=0.4\textwidth]{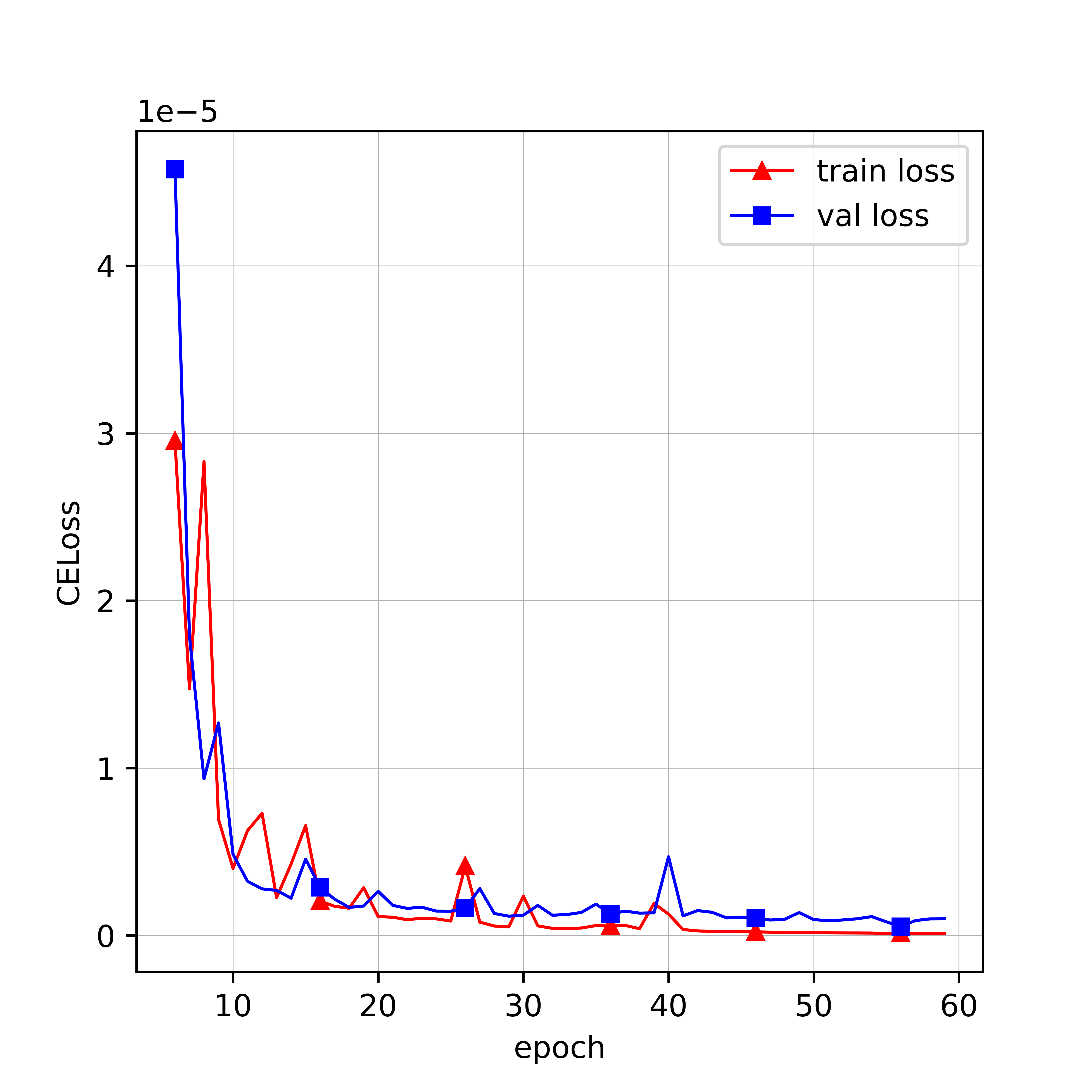}
\caption{The loss curve of the 5GC-former model}
\label{lossg}
\end{figure}

The parameters in training are summarized in Table \ref{tab0}. The loss curves of the models are shown in Fig. \ref{losss} and Fig. \ref{lossg} respectively. The abscissa represents the epoch number while the ordinate represents the loss value of the model defined in \eqref{loss}. As depicted in the figures, the loss value of 5GC-Seq2Seq declines steadily throughout the training, while the loss value of 5GC-former declines with several fluctuations because of the more complex structure, \textcolor{black}{and both of them reach convergence in the later stages of training.} Due to the parallel training, the training time of 5GC-former is 1/12 of that of 5G-Seq2Seq on the same signaling data.

\subsection{Performance Testing and Analysis}
We test the performance of two models on the testing set. \textcolor{black}{Considering the unbalanced distribution of positive and negative samples, we calculate the precision, recall, and F1-score of models.} To further compare the performance of the two models, we also 
specifically test their prediction performance on type {\uppercase\expandafter{\romannumeral 2}} IEs, which are more challenging for DL models to predict.

\begin{table}[!t]
    \centering
    \caption{\textcolor{black}{The performance of models on the test set}}
    \begin{tabular}{cccc}
        \toprule \textbf{Model} & \textbf{Precision} & \textbf{Recall} & \textbf{F1-score} \\
        \midrule \textbf{5GC-Seq2Seq (Overall)} & \textcolor{black}{99.99705\%} & \textcolor{black}{99.99732\%} & \textcolor{black}{99.99719\%}\\
        \textbf{5GC-Former (Overall)} & \textcolor{black}{\textbf{99.99917\%}} & \textcolor{black}{\textbf{99.99917\%}} & \textcolor{black}{\textbf{99.99917\%}} \\
        \textbf{5GC-Seq2Seq (Type {\uppercase\expandafter{\romannumeral 2}})} & \textcolor{black}{99.95568\% }&\textcolor{black}{ 99.95976\%} & \textcolor{black}{99.95771\%}\\
        \textbf{5GC-Former (Type {\uppercase\expandafter{\romannumeral 2}})} & \textcolor{black}{\textbf{99.98785\%}} & \textcolor{black}{\textbf{99.98785\%}} & \textcolor{black}{\textbf{99.98785\%}} \\
        \bottomrule
    \end{tabular}
    \label{tab}
\end{table}

\textcolor{black}{Firstly, we compare the performance of two models on the test set with dialogues generated from a single UE, as shown in Table \ref{tab}. For 5GC-Seq2Seq, the precision is not the same as the recall, indicating instances where the IEs carried in the target signaling message are erroneously predicted as absent or the model predicts uncarried IEs to be present. In contrast, this phenomenon does not exist in 5GC-former.}

\begin{table}[!t]
    \centering
    \caption{\textcolor{black}{The F1-score of models with concurrent UEs}}
    \begin{tabular}{m{1.6cm}ccccc}
        \toprule \textcolor{black}{\textbf{Concurrency number}} & \textcolor{black}{\textbf{1}} & \textcolor{black}{\textbf{2}} & \textcolor{black}{\textbf{3}}& \textcolor{black}{\textbf{4} }& \textcolor{black}{\textbf{5}} \\
        \midrule \textcolor{black}{\textbf{S (Overall)}} & \textcolor{black}{99.997\%} & \textcolor{black}{99.65\%} & \textcolor{black}{99.59\%} & \textcolor{black}{99.34\%} & \textcolor{black}{99.18\%}\\
        \textcolor{black}{\textbf{F (Overall)}} & \textcolor{black}{\textbf{99.999\%}} & \textcolor{black}{\textbf{99.89\%}} & \textcolor{black}{\textbf{99.81\%}} & \textcolor{black}{\textbf{99.72\%}} & \textcolor{black}{\textbf{99.57\%}}\\
        \textcolor{black}{\textbf{S (Type {\uppercase\expandafter{\romannumeral 2}})}}& \textcolor{black}{99.958\%} & \textcolor{black}{96.16\%} & \textcolor{black}{95.27\%} & \textcolor{black}{91.91\%} & \textcolor{black}{89.43\%}\\
        \textcolor{black}{\textbf{F (Type {\uppercase\expandafter{\romannumeral 2}})}} & \textcolor{black}{\textbf{99.988\%}} & \textcolor{black}{\textbf{99.10\% }}& \textcolor{black}{\textbf{98.10\%}} & \textcolor{black}{\textbf{97.04\%}} & \textcolor{black}{\textbf{95.33\%}}  \\
        \bottomrule
    \end{tabular}
    \begin{tablenotes}
        \item[] \textcolor{black}{S and F are the abbreviations for 5GC-Seq2Seq and 5GC-Former.}
    \end{tablenotes}
    \label{tab_con}
\end{table}

\textcolor{black}{Furthermore, we evaluated the performance of the models with varying numbers of concurrent UEs using the F1-score. 
As shown in Table \ref{tab_con}, although the two models perform similarly in the case of a single UE, the performance of 5GC-Seq2Seq deteriorates rapidly as the number of concurrent UEs increases, especially on Type {\uppercase\expandafter{\romannumeral 2}} IEs, while 5GC-former is capable of maintaining a relatively high F1-score. Moreover, as the number of concurrent users increases, the performance gap between the two models widens.}

\textcolor{black}{The experimental results of detection performance are analyzed as follows: 5GC-Seq2Seq, functioning as an RNN-based model, exhibits a natural advantage in handling sequential data within a single procedure as it can continually update the hidden state using the inputs. However, challenges arise when confronted with concurrent procedures and UE dialogues, leading to the interleaving of multiple series of data. In such scenarios, the model needs to update several states asynchronously, which hinders both training and prediction processes. On the other hand, the 5GC-former model, entirely composed of attentional structures, focuses solely on the previous signaling messages that bear the highest relevance to the input. This approach effectively excludes interference from data originating from other procedures or UEs and consequently addresses the aforementioned problem to a certain extent.}
\renewcommand{\arraystretch}{1.4}
\begin{table*}[!t]
    \centering
    \caption{An example for signaling prediction in handover procedure}
    \begin{threeparttable}
    \begin{tabular}{cccc}
    \toprule \textbf{Serial No.} & \textbf{Symbol} &\textbf{Message Direction}  & \textbf{Message type} \\
        \midrule
        1& $s_u^1$ &UE/RAN$\Rightarrow$CP*(AMF)&InitialUEMessage, Registration request\\
2& $s_d^{1,1}$ &CP(AMF)$\Rightarrow$UE/RAN& DownlinkNASTransport, Authentication request\\
3& $s_u^2$ &UE/RAN$\Rightarrow$CP(AMF)&UplinkNASTransport, Authentication response\\
4& $s_d^{2,1}$ &CP(AMF)$\Rightarrow$UE/RAN&DownlinkNASTransport, Security mode command\\
5& $s_u^3$ &UE/RAN$\Rightarrow$CP(AMF)&UplinkNASTransport, Registration request\\
6& $s_d^{3,1}$ &CP(AMF)$\Rightarrow$UE/RAN&InitialContextSetupRequest, Registration accept\\
7& $s_u^4$ &UE/RAN$\Rightarrow$CP(AMF)&InitialContextSetupResponse\\
8& $s_u^5$ &UE/RAN$\Rightarrow$CP(AMF)&UERadioCapabilityInfoIndication\\
9& $s_u^6$ &UE/RAN$\Rightarrow$CP(AMF)&Registration complete\\
10& $s_u^7$ &UE/RAN$\Rightarrow$CP(AMF)&UL NAS transport, PDU session establishment request (1)\\
11& $s_d^{7,1}$ &CP(SMF)$\Rightarrow$UPF&PFCP Session Establishment Request (1)\\
12& $s_u^8$ &UE/RAN$\Rightarrow$CP(AMF)&UL NAS transport, PDU session establishment request (2)\\
13& $s_d^{8,1}$ &CP(SMF)$\Rightarrow$UPF&PFCP Session Establishment Request (2)\\
14& $s_u^9$ &UPF$\Rightarrow$CP(SMF)&PFCP Session Establishment Response (2)\\
15& $s_d^{9,1}$ &CP(AMF)$\Rightarrow$UE/RAN&DL NAS transport, PDU session establishment accept (2)\\
16& $s_u^{10}$ &UE/RAN$\Rightarrow$CP(AMF)&PDU Session Resource Setup Response (2)\\
17& $s_d^{10,1}$ &CP(SMF)$\Rightarrow$UPF&PFCP Session Modification Request (2)\\
18& $s_u^{11}$ &UPF$\Rightarrow$CP(SMF)&PFCP Session Modification Response (2)\\
19& $s_u^{12}$ &UE/RAN$\Rightarrow$CP(AMF)&UL NAS transport, PDU session establishment request (3)\\
20& $s_d^{12,1}$ &CP(SMF)$\Rightarrow$UPF&PFCP Session Establishment Request (3)\\
21& $s_u^{13}$ &UPF$\Rightarrow$CP(SMF)&PFCP Session Establishment Response (3)\\
22& $s_d^{13,1}$ &CP(AMF)$\Rightarrow$UE/RAN&DL NAS transport, PDU session establishment accept (3)\\
23& $s_u^{14}$ &UE/RAN$\Rightarrow$CP(AMF)&PDU Session Resource Setup Response (3)\\
24& $s_d^{14,1}$ &CP(SMF)$\Rightarrow$UPF&PFCP Session Modification Request (3)\\
25& $s_u^{15}$ &UPF$\Rightarrow$CP(SMF)&PFCP Session Modification Response (3)\\
26& $s_u^{16}$ &UPF$\Rightarrow$CP(SMF)&PFCP Session Establishment Response (1)\\
27& $s_d^{16,1}$ &CP(AMF)$\Rightarrow$UE/RAN&DL NAS transport, PDU session establishment accept (1)\\
28& $s_u^{17}$ &UE/RAN$\Rightarrow$CP(AMF)&PDU Session Resource Setup Response (1)\\
29& $s_d^{17,1}$ &CP(SMF)$\Rightarrow$UPF&PFCP Session Modification Request (1)\\
30& $s_u^{18}$ &UPF$\Rightarrow$CP(SMF)&PFCP Session Modification Response (1)\\
31& $s_u^{19}$ &UE/RAN2$\Rightarrow$CP(AMF)&PathSwitchRequest\\
32& $s_d^{19,1}$ &CP(SMF)$\Rightarrow$UPF&PFCP Session Modification Request (1)\\
33& $s_d^{19,2}$ &CP(SMF)$\Rightarrow$UPF&PFCP Session Modification Request (2)\\
34& $s_d^{19,3}$ &CP(SMF)$\Rightarrow$UPF&PFCP Session Modification Request (3)\\
        \bottomrule
    \end{tabular}
    \begin{tablenotes}
		\footnotesize
		\item[*] CP is the abbreviation for Control Plane
    \end{tablenotes}
    \end{threeparttable}
    \label{example}
\end{table*}

\begin{figure*}[!t]
\centering
\includegraphics[width=0.9\textwidth]{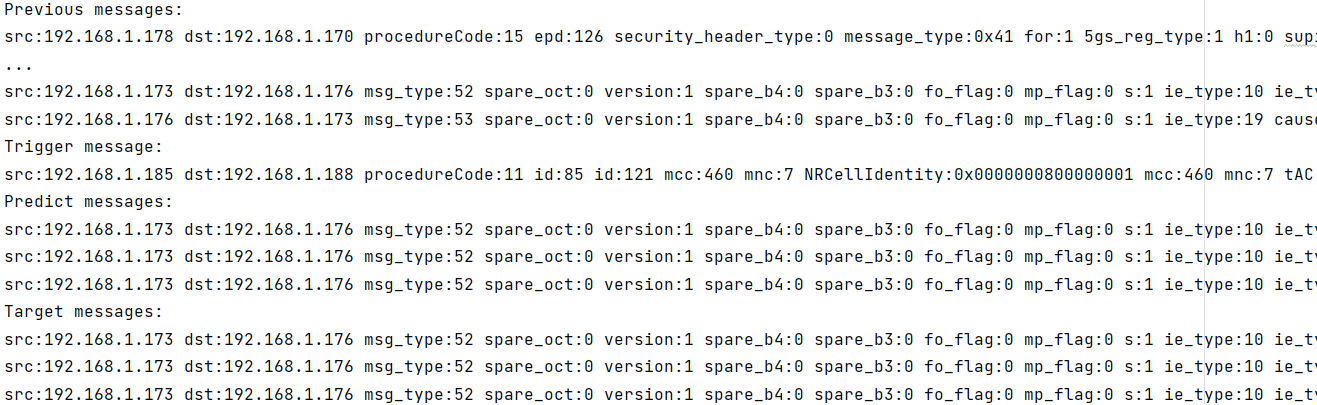}
\caption{The content of predictive messages}
\label{prediction}
\end{figure*}

\begin{figure*}[!t]
\centering
\subfloat[Layer 1]{\includegraphics[width=0.4\textwidth]{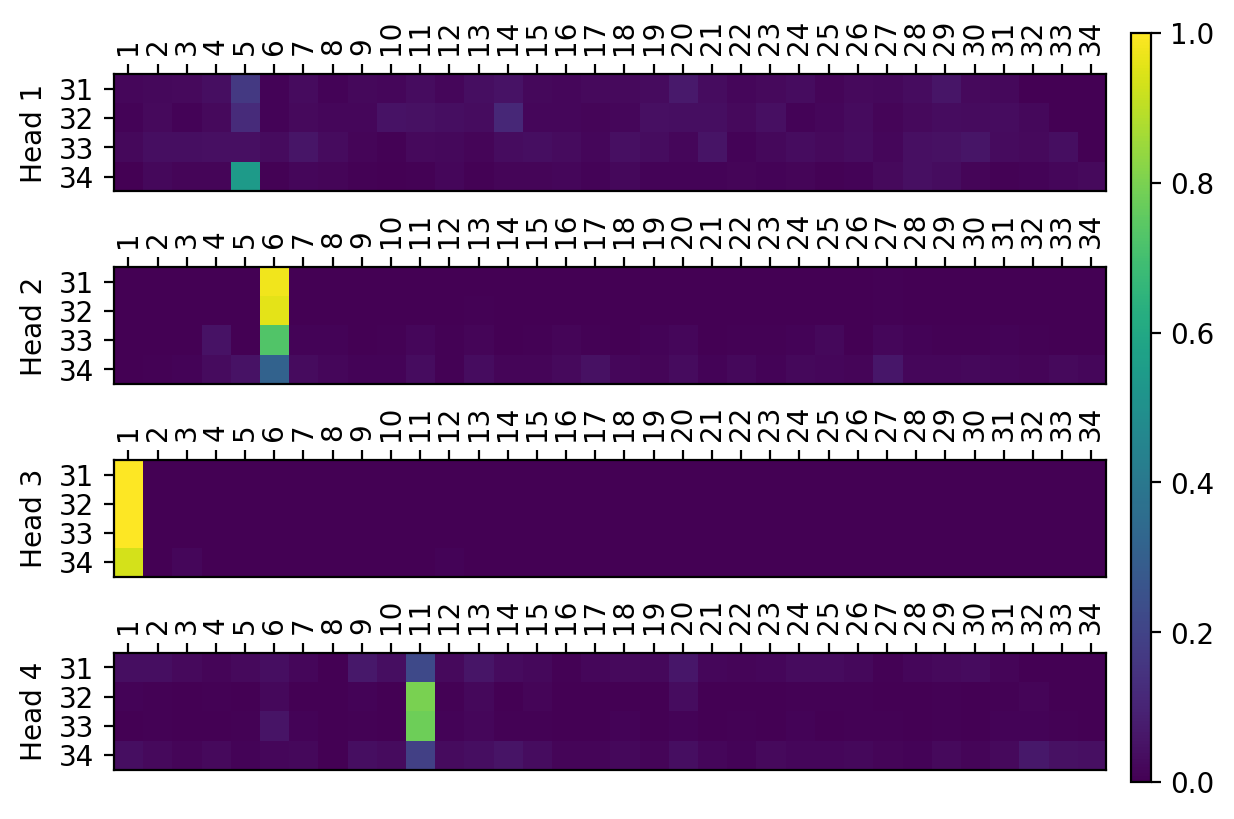}%
\label{l1}
}
\hfil
\subfloat[Layer 2]{\includegraphics[width=0.4\textwidth]{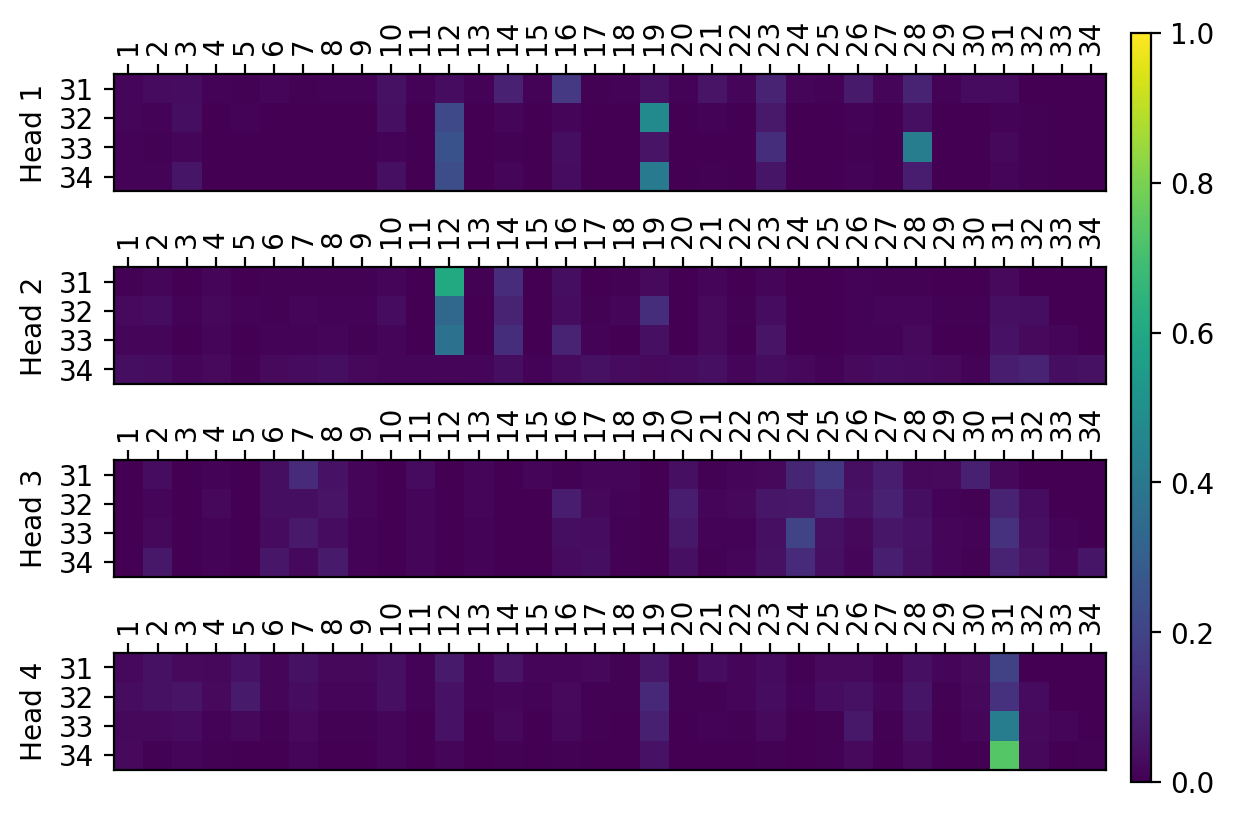}%
\label{l2}}

\subfloat[Layer 3]{\includegraphics[width=0.4\textwidth]{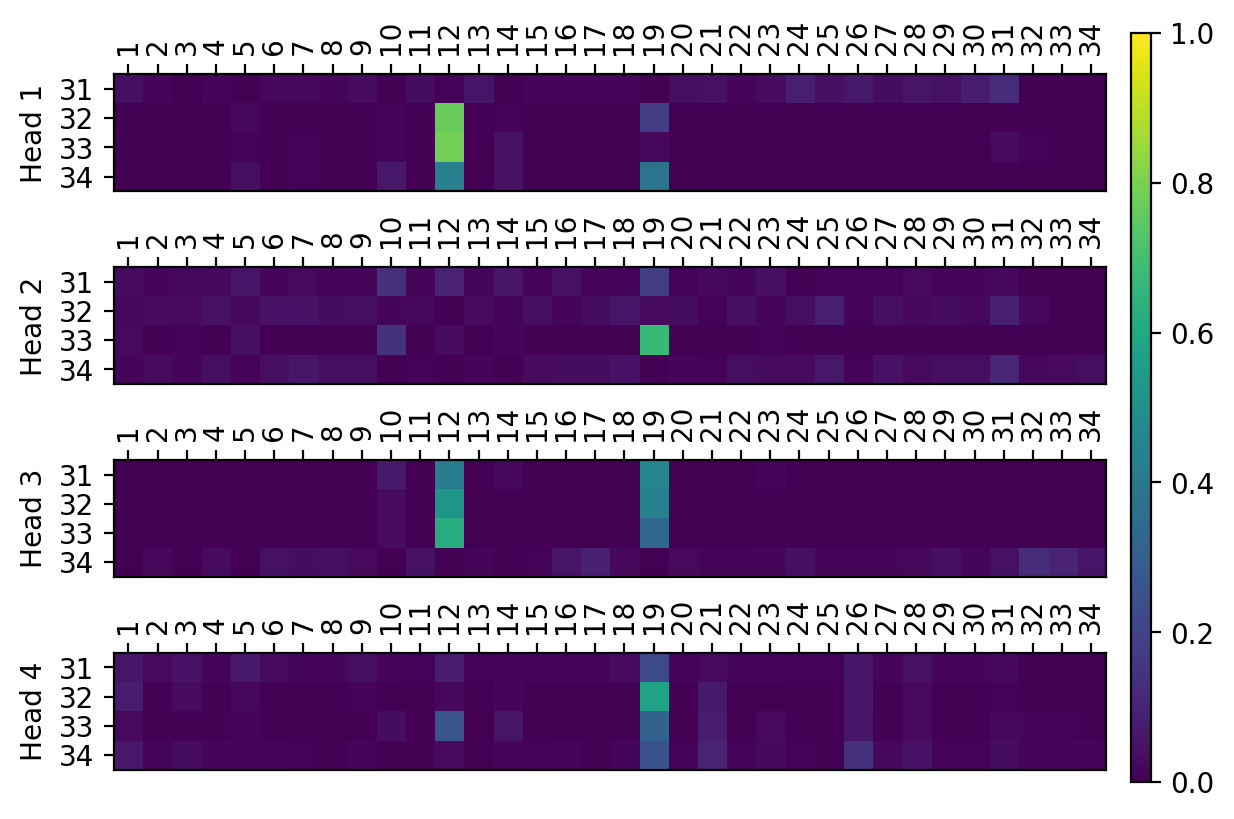}%
\label{l3}}
\hfil
\subfloat[Layer 4]{\includegraphics[width=0.4\textwidth]{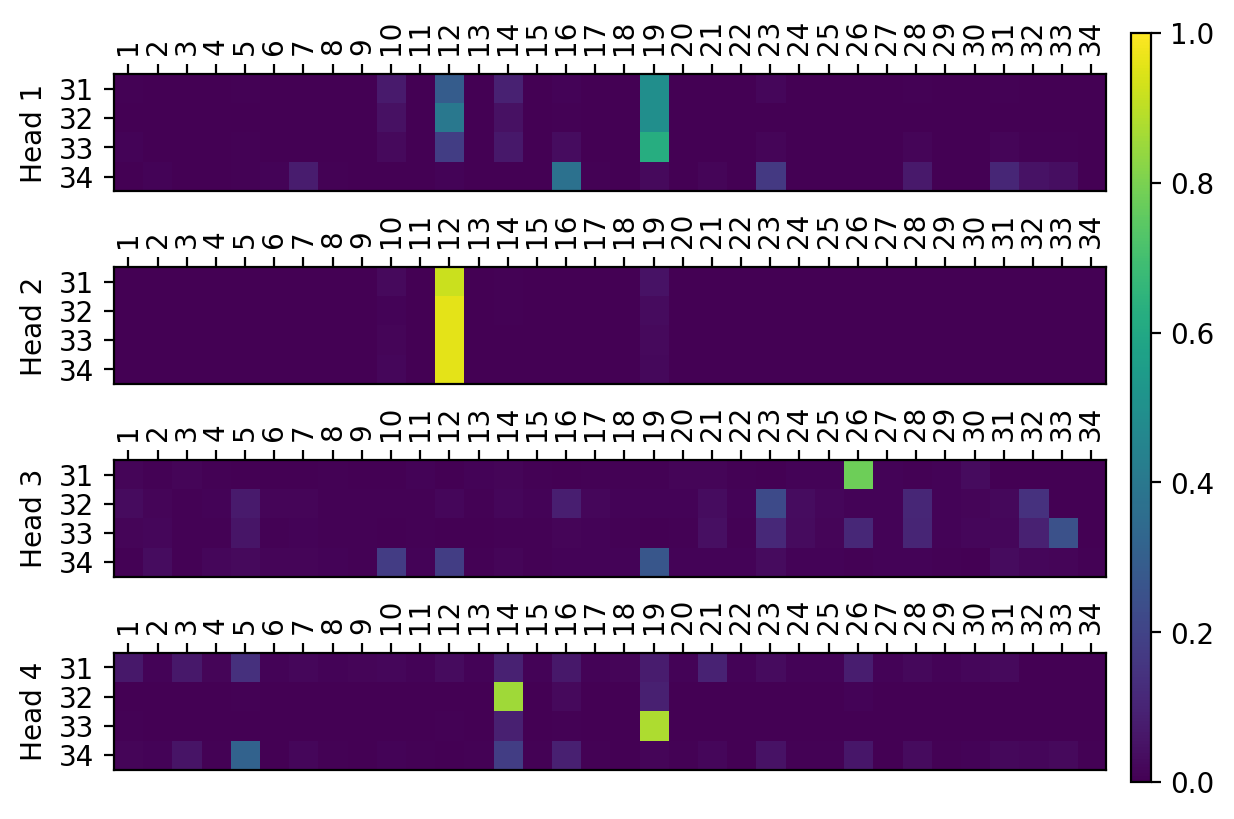}%
\label{l4}}

\subfloat[Layer 5]{\includegraphics[width=0.4\textwidth]{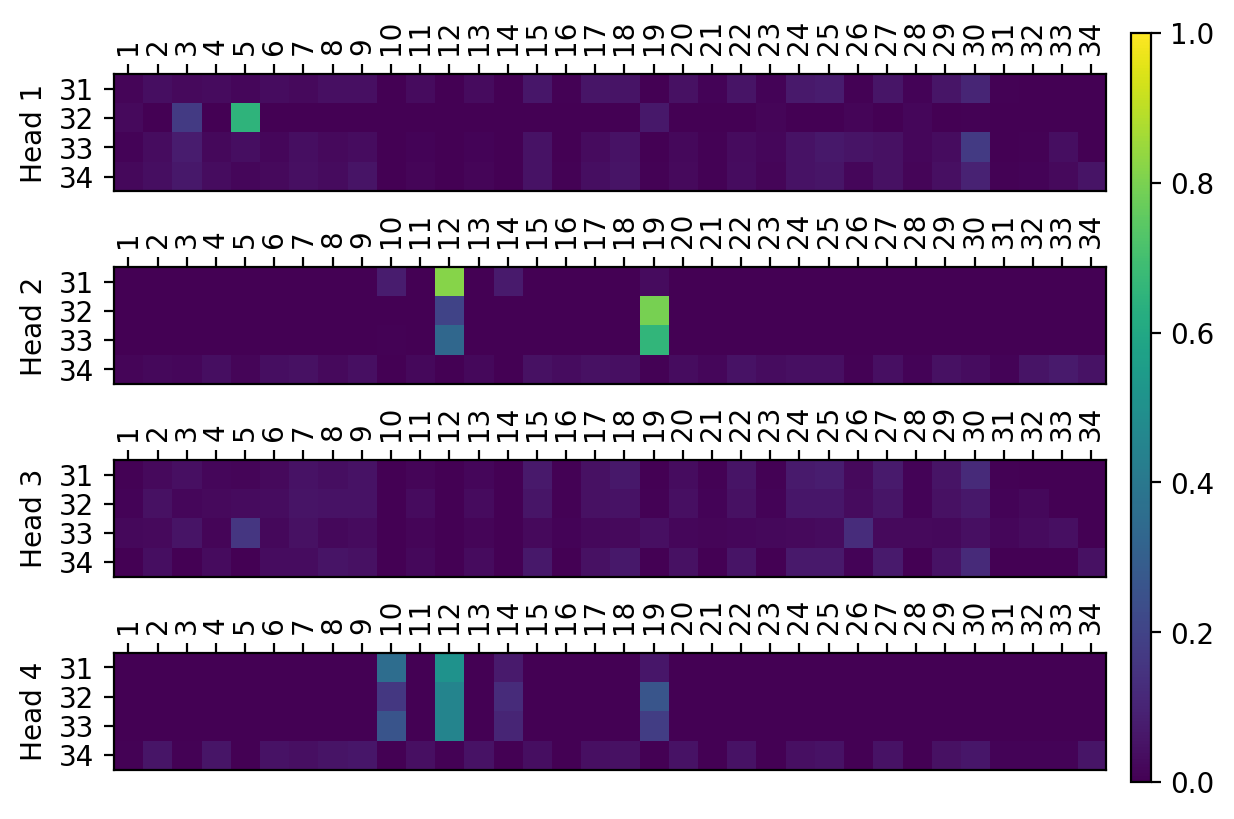}%
\label{l5}}
\hfil
\subfloat[Layer 6]{\includegraphics[width=0.4\textwidth]{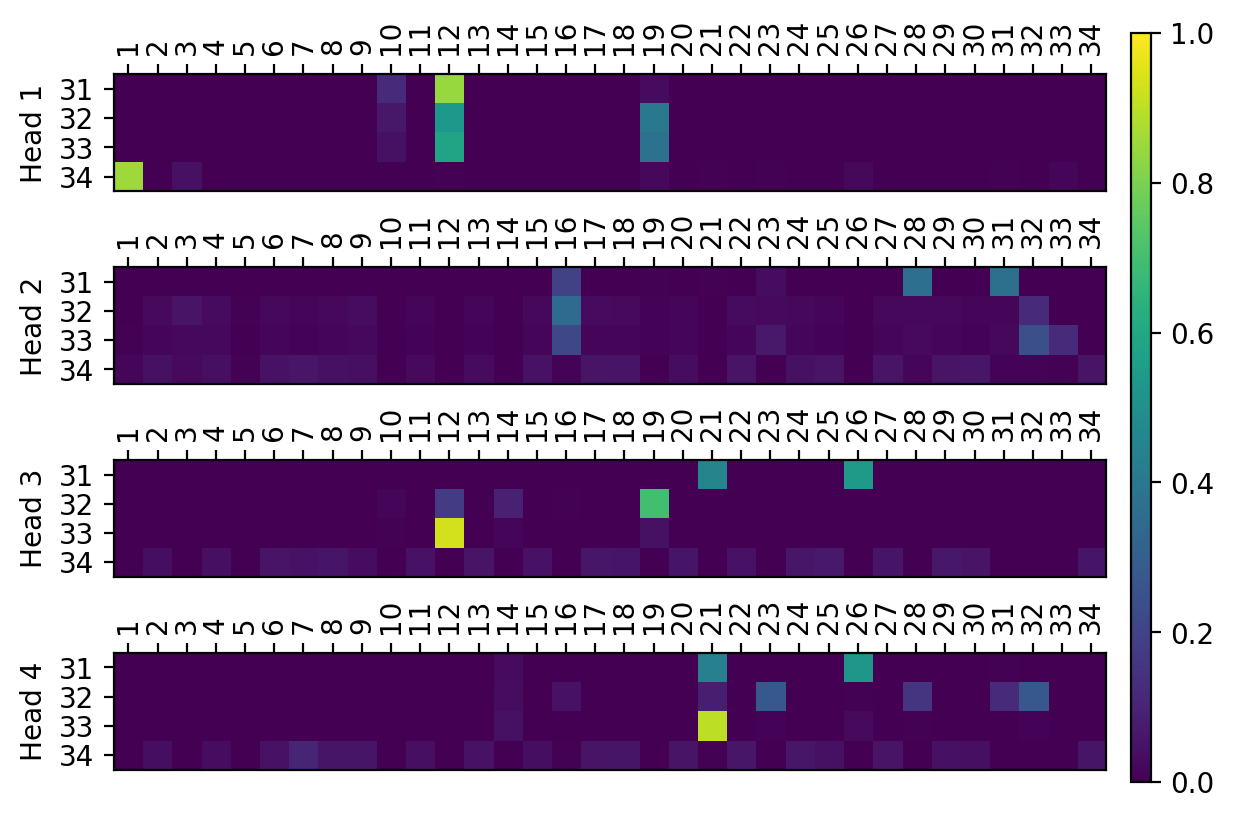}%
\label{l6}}

\caption{Attention Maps}
\label{attention map}
\end{figure*}

An example is also \textcolor{black}{provided} to elaborate on the model predictive process and attention mechanism. As depicted in Table \ref{example}, The UE has \textcolor{black}{successfully} completed the registration procedure and established 3 PDU sessions through 30 signaling messages from $s_u^1$ to $s_u^{18}$. \textcolor{black}{Subsequently,} it sends a handover request $s_u^{19}$ to the control plane from another base station. The trigger message $s_u^{19}$ and previous messages are stacked and fed into the model to get the predictive output $s_d^{19,1}$, then the output is attached to the sequence to perform the next prediction, such \textcolor{black}{process} repeat 4 times until the output becomes $s_{end}$. \textcolor{black}{The content of each message can be translated using dictionaries constructed in Section \ref{dictionary},} as shown in Fig. \ref{prediction}. Due to the excessive length of the whole key, only the last key of the IE tree is presented.

Fig. \ref{attention map} illustrates the attention between messages in different layers and heads \textcolor{black}{by means of heat maps}, from which we can find that in the first layer, the model concentrates on registration-related messages such as the 1st message (\textit{registration\_request}) and the 6th message (\textit{registration\_request}), while in higher layers, the model pays more attention to PDU session-related messages, including the 10th, 12th, 19th messages (\textit{PDU\_session\_establishment\_request}) and nearby messages. It is worth noting that as the layers become higher, the model's attention \textcolor{black}{tends to disperse}, facilitating the consideration of overall information. Meanwhile, \textcolor{black}{due to the residual connections, certain heads continue to focus on specific information,} such as head 2 of layer 4, head 2, and head 4 of layer 5, etc. The \textcolor{black}{effective} coordination between specific and global attention allows for better performance of the prediction model. The unidirectionality of the 5G-former is also depicted in Fig. \ref{attention map}. It can be noticed that the top right corner of each attention map is black, particularly noticeable in head 3 of layer 4 and head 2 of layer 6, indicating that each message only pays attention to itself and the previous messages.

\textcolor{black}{By analyzing the structure and performance of the two models, we can deduce their advantages and disadvantages respectively. The 5GC-seq2seq model has a relatively simple structure, low computational complexity, and storage efficiency as it utilizes encoder-side state updates to pass information. However, its predictive performance in concurrent UE scenarios is unsatisfactory, and substantial training time is necessary due to the absence of parallelism. The 5GC-former model, on the other hand, has better predictive performance, reduced training time, and the potential to construct larger models. Nevertheless, it necessitates the reservation of previous messages and is computationally complex, making it unsuitable for scenarios requiring extremely prompt responses.}

\section{Conclusion}
In this paper, we propose two novel data-driven architectures for modeling the behavior of the 5G control plane. Specifically, we implement two deep learning models, 5GC-Seq2Seq and 5GC-former, based on the Vanilla Seq2Seq model and Transformer decoder respectively. We also design a solution that allows the signaling messages to be interconverted with \textcolor{black}{length-limited} vectors and construct datasets on signaling messages from various procedures generated by the Spirent C50 network tester. 
\textcolor{black}{Our results show that in the single UE scenario, 5GC-Seq2Seq achieves an F1-score of 99.997\%  with a simple structure, but exhibits poor performance in handling concurrency. In contrast, 5GC-former attains an F1-score exceeding 99.999\% and maintains a relatively high F1-score even during multi-UE concurrency by constructing a more intricate and highly parallel model.}

\textcolor{black}{In future studies, we will explore a swifter and more precise predictive approach by employing enhanced structure and innovative cells within adapted DL models. In addition, to deal with the labeling requirement in the current scheme for specific IEs associated with prior knowledge beyond neural network capabilities, such as authentication or ID allocation, we should consider efficient solutions to establish connections between DL models and databases within the 5G core. Transfer learning for different physical networks, which can significantly reduce the training time on new networks, should be analyzed as well.}

%



\section*{Acknowledgment}
This work was supported in part by the National Key R\&D Program under Grant 2020YFB1806600 and 2022YFB2901601, as well as the Major Key Project of Peng Cheng Laboratory under Grant PCL 2021A01-2.

\ifCLASSOPTIONcaptionsoff
  \newpage
\fi



\bibliographystyle{IEEEtranN}

\bibliography{IEEEabrv,IEEEexample}
\end{document}